\documentstyle[amsmath,amssymb,graphicx]{article}

\def\be{\begin{eqnarray}}
\def\ee{\end{eqnarray}}
\def\nn{\nonumber}

\hoffset= - 1.0in         

\title{{\bf Conformal blocks as Dotsenko-Fateev Integral
Discriminants} \vspace{.2cm}}
\author{{\bf A.Mironov}\footnote{ {\small {\it
Lebedev Physics Institute} and {\it ITEP, Moscow, Russia}};
mironov@itep.ru; mironov@lpi.ru}, {\bf A.Morozov}\thanks{{\small
{\it ITEP, Moscow, Russia}}; morozov@itep.ru} \ and {\bf
Sh.Shakirov}\thanks{{\small {\it ITEP, Moscow, Russia} and {\it
MIPT, Dolgoprudny, Russia}}; shakirov@itep.ru}\date{ }}

\begin{document}

\maketitle

\vspace{-6.0cm}

\begin{center}
\hfill FIAN/TD-01/10\\
\hfill ITEP/TH-01/10\\
\end{center}

\vspace{4cm}

\begin{abstract}
As anticipated in \cite{DVagt}, elaborated in \cite{Ito,Egu,Wilma},
and explicitly formulated in \cite{MMSh1}, the Dotsenko-Fateev
integral discriminant coincides with conformal blocks, thus
providing an elegant approach to the AGT conjecture, without any reference
to an auxiliary subject of Nekrasov functions. Internal dimensions of
conformal blocks in this identification are associated with the
choice of contours: parameters of the DV phase of the corresponding
matrix models. In this paper we provide further evidence in support
of this identity for the 6-parametric family of the 4-point
spherical conformal blocks, up to level 3 and for arbitrary values
of external dimensions and central charges. We also extend this
result to multi-point spherical functions and comment on a similar
description of the 1-point function on a torus.
\end{abstract}

\bigskip

\bigskip

\tableofcontents

\section{Introduction}

Non-Gaussian integrals are slowly attracting a growing attention,
and start to play a role in modern theoretical studies.
For quantum field theory they have an especially important property:
non-Gaussian integrals are not fully defined by the action,
they also depend on the discrete choice of integration contour,
and this freedom itself depends on the shape (degree) of the action.
Accordingly, the Ward identities,
reflecting the freedom to change integration variables
and describing the coupling constant (RG) dependence of the integral
\cite{UFN},
in non-Gaussian case have several solutions, in one-to-one
correspondence with the possible choice of integration contours.
The closest subjects in pure mathematics are
Picard-Fucks equations
and emerging theory of motivic integration \cite{mot},
and they are actively used, say, in Seiberg-Witten theory \cite{SW}-\cite{SWrev}
and closely related string models.
Initial steps in a more direct approach to the study of non-Gaussian integrals
were recently described in \cite{MShID}, where emphasize is put
on their similarity to the ordinary resultant theory.
Following those papers we use the term {\it integral discriminants}
for non-Gaussian integrals considered from this perspective.
There is of course a parallel development in the field of matrix models,
where above-mentioned ambiguity (contour dependence) could not
remain unnoticed.
There is a long chain of matrix model papers, devoted to this issue,
which finally culminated in the theory of Dijkgraaf-Vafa phases
\cite{DVph} and check-operators \cite{checkop}.

A new boost to a variety of prominent research directions
was recently given by the AGT conjecture \cite{AGT1}-\cite{AGTl},\cite{DVagt}-\cite{MMSh1},
which unified them all in a single entity and allowed to study
the problems of one direction by means of the others.
The present paper is devoted to a particular study of this type:
representation of conformal blocks in terms of the matrix model integral
discriminants with the Dotsenko-Fateev action
(also known as conformal matrix models \cite{KMMMP,UFN} and $\beta$-ensembles
\cite{betas}),
which was suggested in \cite{AGT1,MMnf,DVagt}, further investigated in
\cite{Ito,Egu,Wilma} and finally put into a clear and explicit form
in \cite{MMSh1}.
As emphasized in \cite{MMnf,MMSh1} this also resolves the old
puzzle in conformal field theory (CFT) \cite{CFT}:
how arbitrary conformal blocks are described in terms
of the Dotsenko-Fateev free field correlators \cite{DF,GMMOS}.
The problem was that the number of free parameters in conformal
block (dimensions of internal and external lines) exceeds the number
of parameters in the free field conformal theory (there is no
obvious room for internal-line dimensions and also the choice of
external dimensions is constrained by the peculiar free field conservation
law $\sum\vec\alpha_i = \vec Q$).
As we now know from \cite{DVagt} and \cite{MMSh1},
the lacking parameters are exactly
those of the DV phases, which parameterize the choice of contour of the
eigenvalue integration in matrix integral, i.e. the choice of
contours in the Dotsenko-Fateev screening operators,
with conservation condition omitted.
This was of course always expected in CFT,
the change is that now one has a clear and unambiguous description
of the phenomenon.
In this paper we provide further evidence and further details
about this description, by checking this new version of the AGT relation,
between conformal blocks and Dotsenko-Fateev integral discriminants,
at more levels and for more dimensions than in \cite{MMSh1}.
A proof of this relation, which contains no reference to the still
mysterious Nekrasov functions
(though some clarity seems to emerge here as well, see \cite{NSh,MMBS,MMSh1}),
and should be, therefore, self-contained and straightforward,
remains beyond the scope of the present paper.

\section{Conformal block as Dotsenko-Fateev integral \cite{MMSh1}}

In conformal field theory, the simplest correlator
with non-trivial coordinate dependence is the 4-point correlator
\begin{align*}
\Big< V_{\Delta_1}(z_1, {\bar z}_1) V_{\Delta_2}(z_2,{\bar z}_2)
V_{\Delta_3}(z_3,{\bar z}_3) V_{\Delta_4}(z_4, {\bar z}_4) \Big>
\end{align*}
where $\Delta_i$ are the dimensions of 4 primary fields on a sphere.
With $SL(2)$ transformations $z \mapsto \frac{a z + b}{c z + d}$ with
$ad-bc = 1$, one can always put the four coordinates to $0,q, 1$ and
$\infty$. Hence the correlator depends (up to a simple factor)
on a single variable $q =
\frac{(z_1 - z_2)(z_3-z_4)}{(z_1 - z_3)(z_2 - z_4)}$. By usual CFT
arguments, this 4-point correlator can be written as a sum over a
single intermediate dimension $\Delta$ of a product of the
holomorphic and anti-holomorphic conformal blocks:
\be
\Big<
V_{\Delta_1}(0,0) V_{\Delta_2}(q, {\bar q}) V_{\Delta_3}(1,1)
V_{\Delta_4}(\infty,\infty) \Big> = \sum\limits_{\Delta} \ C \big(
\Delta_1, \Delta_2, \Delta \big) C \big( \Delta, \Delta_3, \Delta_4
\big) \ \times \ {\cal F}\big( \Delta_1, \Delta_2, \Delta_3,
\Delta_4, \Delta, c \, \big| \ q \big) \ee with a shorthand notation
\be {\cal F}\big( \Delta_1, \Delta_2, \Delta_3, \Delta_4, \Delta, c
\, \big| \ q \big) = q^{\Delta - \Delta_1 - \Delta_2} \
q^{\bar\Delta - \bar\Delta_1 - \bar\Delta_2} \ B\big( \Delta_1,
\Delta_2, \Delta_3, \Delta_4, \Delta, c \, \big| \ q \big) \
{\overline B}\big( \bar\Delta_1, \bar\Delta_2, \bar\Delta_3,
\bar\Delta_4, \bar\Delta, c \, \big| \ {\bar q} \big)
\ee
In the r.h.s., $C$'s are the 3-point functions
(they do not depend on $q$ and play the role of normalization
constants needed to make $B(0) = 1$), $c$ is the central charge and
$B(q)$ is the 4-point conformal block. Note that
the bar here means not the complex conjugation,
but the parameters of the anti-holomorphic conformal
block, see \cite{CFT} for more details.

The function $B(q)$ is now widely recognized as the simplest representative of a family
of important special functions of string theory,
which appear not only in $2d$ conformal field theory,
but also in $4d$ supersymmetric field theory,
according to the AGT conjecture \cite{AGT1}.
These special functions generalize in a clever way the hypergeometric functions \cite{MMnf},
and should of course possess various complementary representations,
the most important one being \emph{series} and \emph{integral} representations.
The latter one constitutes the subject of our paper.

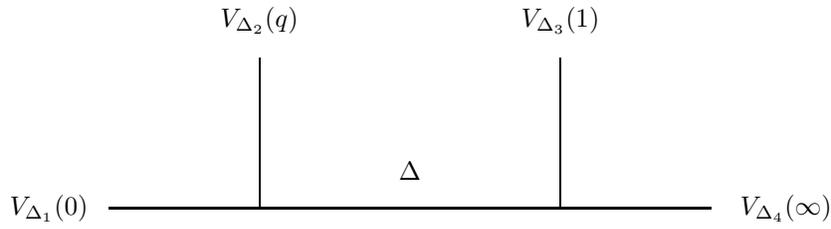
\begin{figure}\begin{center}
\unitlength 1mm 
\linethickness{0.4pt}
\ifx\plotpoint\undefined\newsavebox{\plotpoint}\fi 
\begin{picture}(100,20)(0,8)
%
%
\put(10,10){\line(1,0){80}}
\put(30,10){\line(0,1){20}}
\put(70,10){\line(0,1){20}}
\put(2,10){\makebox(0,0)[cc]{$V_{\Delta_1}(0)$}}
\put(30,35){\makebox(0,0)[cc]{$V_{\Delta_2}(q)$}}
\put(70,35){\makebox(0,0)[cc]{$V_{\Delta_3}(1)$}}
\put(100,10){\makebox(0,0)[cc]{$V_{\Delta_4}(\infty)$}}
\put(50,15){\makebox(0,0)[cc]{$\Delta$}}
\end{picture}
\caption{\footnotesize{
Feynman-like diagram for a 4-point conformal block.
}}\label{4block}
\end{center}\end{figure}

Historically, a series representation for $B(q)$ was found the first.
It is obtained by the decomposition of correlators via iterating the operator product
expansions.
This procedure is extensively reviewed in the literature \cite{MMMagt, MMnf, AlAnd2},
and, in the 4-point case shown in Fig.\ref{4block},
it gives the following series expansion in powers of $q$
\be
B\Big( \Delta_1, \Delta_2, \Delta_3, \Delta_4, \Delta, c \ \Big| \ q \Big)
= \sum\limits_{|Y| = |Y^{\prime}|} \ q^{|Y|} \
\gamma_{\Delta_1 \Delta_2 \Delta}(Y) Q_{\Delta}^{-1}(Y,Y^{\prime})
\gamma_{\Delta \Delta_3 \Delta_4}(Y^{\prime})
\label{B4Series}
\ee
where the sum goes over the Young diagrams $Y = (k_1 \geq k_2 \geq \ldots)$ and
$Y^{\prime} = (k^{\prime}_1 \geq k^{\prime}_2 \geq \ldots)$ of equal size
$|Y| = k_1 + k_2 + \ldots = k^{\prime}_1 + k^{\prime}_2 + \ldots = |Y^{\prime}|$,
parameterizing the Virasoro descendants in the intermediate channel.
The relevant values of the Virasoro triple vertices (the structure constants of operator
product expansion) are given by
\be
\gamma_{\Delta_1 \Delta_2 \Delta_3}(Y) = \prod\limits_{i}
\big( k_1 \Delta_1 + \Delta_3 - \Delta_2 + k_1 + \ldots + k_{i-1} \big)
\ee
and
\be
 Q_\Delta(Y,Y') =\  < \Delta | L_{Y} L_{-Y^{\prime}} | \Delta >\ =
\label{table}
\ee
\be
\begin{array}{|c||c||c||c|c||c|c|c||c|}
\hline
Y/Y' & \emptyset & [1] & [2] & [11] & [3] & [21] & [111] &\ldots\\
\hline\hline
\emptyset &0&&&&&&&\\
\hline\hline
[1] &&2\Delta&&&&&&\\
\hline\hline
[2] &&&\frac{1}{2}(8\Delta + c)&6\Delta&&&&\\
\hline
[11] &&&6\Delta&4\Delta(1+2\Delta)&&&&\\
\hline\hline
[3] &&&&&6\Delta+2c&2(8\Delta+c)&24\Delta&\\
\hline
[21] &&&&&2(8\Delta+c)&8\Delta^2+(34+c)\Delta + 2c&36\Delta(\Delta+1)&\\
\hline
[111] &&&&&24\Delta&36\Delta(\Delta+1)&24\Delta(\Delta+1)(2\Delta+1)&\\
\hline\hline
\ldots &&&&&&&&\\
\hline
\end{array}
\nn
\ee
is the Shapovalov matrix of the Virasoro algebra.
Both $\gamma$ and $Q^{-1}$ are straightforwardly calculable,
what makes (\ref{B4Series}) an explicit and useful series expansion.

Less investigated is the second chapter of the reference book of special functions, that is,
\emph{integral} representation of the conformal blocks.
Such a representation was actually proposed by
V.Dotsenko and V.Fateev \cite{DF} (see also \cite{GMMOS})
in terms of the free field correlators
\be
\left<\left< :e^{\tilde\alpha_1\phi(z_1)}: \ \ldots \
:e^{\tilde\alpha_m\phi(z_m)}: \right>\right> =
\prod\limits_{ 1 \leq i < j \leq m} (z_j - z_i)^{2\tilde\alpha_i \tilde\alpha_j}
\ee
integrated over \emph{a part of} the variables $z_i$ with \emph{some}
choice of contours $C_i$.
Integrated can be only operators of unit dimensions, i.e. the corresponding
$\tilde\alpha$ are fixed to particular values, called $b$ or $-1/b$.
Integrated operators are often called "screening charges" or simply "screenings".
Generically, the screenings of only one type (say, $b$) are involved
\cite{KMMMP},
though in rational conformal models the both do essentially contribute \cite{DF}.
The precise choice of integration contours remained a mystery for quite a long time,
until the recent breakthrough \cite{DVagt}-\cite{MMSh1},
motivated by the AGT conjecture \cite{AGT1}.
Following \cite{MMSh1}, we make a very simple choice of contours for the
Dotsenko-Fateev partition function
\begin{align}
Z_{DF} = \left<\left< :e^{\tilde\alpha_1\phi(0)}:\ :e^{\tilde\alpha_2\phi(q)}:\
:e^{\tilde\alpha_3\phi(1)}:\ :e^{\tilde\alpha_4\phi(\infty)}:\
\left(\int_0^q :e^{b\phi(z)}:\,dz\right)^{N_1}
\left(\int_0^1 :e^{b\phi(z)}:\,dz\right)^{N_2}\right>\right> = \nn \\
\boxed{ \ \ \ = q^{\dfrac{\alpha_1 \alpha_2}{2 \beta}} \
(1 - q)^{\dfrac{\alpha_2 \alpha_3}{2 \beta}}
\prod\limits_{i = 1}^{N_1} \int\limits_{0}^{q} d z_i \
\prod\limits_{i = N_1+1}^{N_1 + N_2} \int\limits_{0}^{1} d z_i \
\prod\limits_{i < j} (z_j - z_i)^{2 \beta}
\prod\limits_{i} z_i^{\alpha_1} (z_i - q)^{\alpha_2} (z_i - 1)^{\alpha_3} \ \ \ }
\label{FD4pt}
\end{align}
where $\beta = b^2$ and $\alpha_i = 2 b \tilde\alpha_i$.
Our main statement, which we check in detail,
is that {\bf the Dotsenko-Fateev integral with this simple choice
of integration contours precisely reproduces the 4-point conformal block:}
\begin{align}
\boxed{
Z_{DF}\Big( \alpha_1, \alpha_2, \alpha_3, N_1, N_2, \beta \ \Big| \ q \ \Big) =
C_{DF} \cdot q^{\Delta - \Delta_1 - \Delta_2} \cdot
B\Big( \Delta_1, \Delta_2, \Delta_3, \Delta_4, \Delta, c \ \Big| \ q \Big)}
\label{MainEquality}
\end{align}
where $C_{DF}$ is the Dotsenko-Fateev normalization constant,
which does not depend on $q$ (but depends on all the other parameters).

\begin{figure}\begin{center}
\unitlength 1mm 
\linethickness{0.4pt}
\ifx\plotpoint\undefined\newsavebox{\plotpoint}\fi 
\begin{picture}(100,20)(0,8)
%
%
\put(10,10){\line(1,0){80}}
\put(30,10){\circle{4}}
\put(30,10){\line(0,1){20}}
\put(30,4){\makebox(0,0)[cc]{$N_1$}}
\put(70,10){\circle{4}}
\put(70,10){\line(0,1){20}}
\put(70,4){\makebox(0,0)[cc]{$N_2$}}
\put(5,15){\makebox(0,0)[cc]{$V_{\Delta_1}(0)$}}
\put(30,35){\makebox(0,0)[cc]{$V_{\Delta_2}(q)$}}
\put(70,35){\makebox(0,0)[cc]{$V_{\Delta_3}(1)$}}
\put(95,15){\makebox(0,0)[cc]{$V_{\Delta_4}(\infty)$}}
\put(50,15){\makebox(0,0)[cc]{$\Delta$}}
\put(15,6){\makebox(0,0){$\tilde \alpha_1$}}
\put(35,24){\makebox(0,0){$\tilde \alpha_2$}}
\put(75,24){\makebox(0,0){$\tilde \alpha_3$}}
\put(50,6){\makebox(0,0){$\tilde a = \tilde\alpha_1+\tilde\alpha_2 + N_1b$}}
\put(95,6){\makebox(0,0)[cc]{$\tilde\alpha_4\cong\tilde a + \tilde\alpha_3+N_2b$}}
\end{picture}
\vspace{3ex}
\caption{\footnotesize{
Feynman-like diagram for a 4-point conformal block.
Here $\Delta = \tilde a \left( \tilde a + \frac{1}{b}-b\right)$ and
$\Delta_j = \tilde\alpha_j\left(\tilde\alpha_j + \frac{1}{b}-b\right)$
are the single internal and four external dimensions, respectively.
The role of the screenings is to modify the free field selection rule at the vertices:
instead of $\tilde a = \tilde\alpha_1 + \tilde\alpha_2$ and
$\tilde\alpha_4 \cong \tilde a + \alpha_3$ one has
$\tilde a = \tilde\alpha_1 + \tilde\alpha_2 + b N_1$ and $\tilde\alpha_4 =
b-1/b-\tilde\alpha_1-\tilde\alpha_2-\tilde\alpha_3-b(N_1+N_2)
\cong \tilde a + \tilde\alpha_3 + b N_2$.
}}\label{4blockCircles}
\end{center}\end{figure}
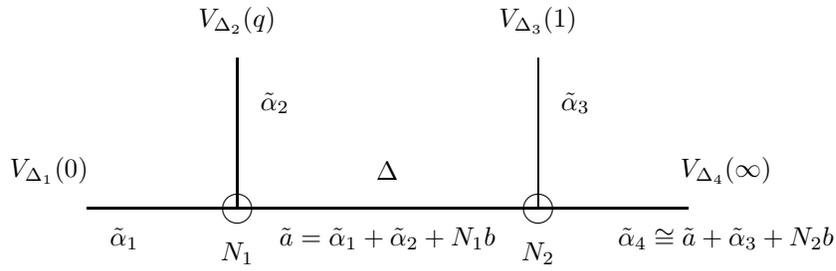

The aim of present paper is to study relation (\ref{MainEquality})
more thoroughly and find an explicit correspondence between the six parameters
of the conformal block $\Delta_1, \Delta_2, \Delta_3, \Delta_4, \Delta, c$,
and the six parameters of the Dotsenko-Fateev integral
$\alpha_1,\alpha_2,\alpha_3,N_1,N_2, \beta$.
Our result in the 4-point sector is
\begin{align}
\left\{
\begin{array}{lllll}
\Delta_1 = \dfrac{ \alpha_1 ( \alpha_1 + 2 - 2 \beta )}{4 \beta}, \ \
\Delta_2 = \dfrac{ \alpha_2 ( \alpha_2 + 2 - 2 \beta )}{4 \beta}, \ \
\Delta_3 = \dfrac{ \alpha_3 ( \alpha_3 + 2 - 2 \beta )}{4 \beta} \\
\\
\Delta_4 = \dfrac{(2 \beta (N_1 + N_2 ) + \alpha_1 + \alpha_2 + \alpha_3 )
(2 \beta (N_1 + N_2) + \alpha_1 + \alpha_2 + \alpha_3 + 2-2 \beta)}{4\beta} \\
\\
\Delta = \dfrac{(2 \beta N_1 + \alpha_1 + \alpha_2 )
(2 \beta N_1 + \alpha_1 + \alpha_2 + 2 - 2 \beta )}{4 \beta}
\\
\\
c = 1 - 6 \left( \sqrt{\beta} - \dfrac{1}{\sqrt{\beta}} \right)^2 \
\\
\end{array}
\right.
\label{Relations1}
\end{align}
or in terms of the free field variables
\begin{align}
\left\{
\begin{array}{lllll}
\Delta_1 = \tilde\alpha_1 \left( \tilde\alpha_1 + \dfrac{1}{b} - b \right), \ \
\Delta_2 = \tilde\alpha_2 \left( \tilde\alpha_2 + \dfrac{1}{b} - b \right), \ \
\Delta_3 = \tilde\alpha_3 \left( \tilde\alpha_3 + \dfrac{1}{b} - b \right) \\
\\
\Delta_4 = \Big( b (N_1 + N_2 ) + \tilde\alpha_1 + \tilde\alpha_2 +
\tilde\alpha_3 \Big) \left(b (N_1 + N_2) + \tilde\alpha_1 + \tilde\alpha_2
+ \tilde\alpha_3 + \dfrac{1}{b} - b \right) \\
\\
\Delta = \Big( b N_1 + \tilde\alpha_1 + \tilde\alpha_2 \Big)
 \left(b N_1 + \tilde\alpha_1 + \tilde\alpha_2 + \dfrac{1}{b} - b \right)
\\
\\
c = 1 - 6 \left( b - \dfrac{1}{b} \right)^2 \ \\
\end{array}
\right.
\label{Relations2}
\end{align}
Clearly the rules are simple, see Fig.\ref{4blockCircles}:

{\bf 1) Insertion of $N_1$ screening integrals is needed to satisfy the free field
conservation law for the first (left) vertex,
$\tilde a = \tilde\alpha_1+\tilde\alpha_2 + bN_1$ so that the internal dimension
$\Delta = \tilde a\left(\tilde a + \frac{1}{b}-b\right)$
becomes unrelated to the free field value $\Delta_{free}=
(\tilde\alpha_1+\tilde\alpha_2)((\tilde\alpha_1+\tilde\alpha_2+{1\over b}-b)$.
Integrals in these $N_1$ screenings are around positions of $V_{\Delta_1}(0)$
and $V_{\Delta_2}(q)$.}

{\bf 2) Insertion of $N_2$ screening integrals is needed to satisfy the free field
conservation law for the second (right) vertex,
$\tilde\alpha_4\cong \tilde a+\tilde\alpha_3 + bN_2$.
}

{\bf 3) Alternatively, these additional $N_2$ screening integrals are needed to satisfy
the free field conservation law
\be
\sum_{i=1}^4\tilde\alpha_i + b(N_1+N_2) = (g-1)\left(\frac{1}{b}-b\right)
\ee
($g$ is the genus, $g=0$ for the sphere),
by putting
$\tilde\alpha_4 = -\tilde \alpha_1-\tilde\alpha_2-\tilde\alpha_3
- b(N_1+N_2) - \frac{1}{b} + b$.}

Note that, in the free field formalism, there are two different $\tilde\alpha$-parameters
associated with a given dimension: $\Delta[\tilde\alpha ]=\Delta[b-{1\over b}-\tilde\alpha]$.
We denote this reflection in the $\tilde\alpha$-space by $\cong$: $\tilde\alpha\cong
b-{1\over b}-\tilde\alpha$.

\bigskip

We derive relations (\ref{Relations2}) by expanding the both sides of the main relation
(\ref{MainEquality}) into power series
\begin{align}
Z_{DF}\Big(\alpha_1, \alpha_2, \alpha_3, N_1, N_2, \beta \ \Big| \ q \ \Big)
= C_{DF} \cdot q^{{\rm deg}_{DF}} \cdot
\left[ \ 1 + \sum\limits_{k = 1}^{\infty} \ q^k \!
J_k \Big(\alpha_1, \alpha_2, \alpha_3, N_1, N_2, \beta \Big)  \right]
\label{Expansion1}
\end{align}
\begin{align}
B\Big( \Delta_1, \Delta_2, \Delta_3, \Delta_4, \Delta, c \ \Big| \ q \ \Big) =
1 + \sum\limits_{k = 1}^{\infty} q^k
B_k\Big( \ \Delta_1, \Delta_2, \Delta_3, \Delta_4, \Delta, c \Big)
\label{Expansion2}
\end{align}
\smallskip\\
and comparing the newly derived coefficients $J_k$
with the known coefficients $B_k$ (see, for example, \cite{MMMagt}):
\begin{align*}
B_1\Big( \Delta_1, \Delta_2, \Delta_3, \Delta_4, \Delta, c \Big) =
{(\Delta+\Delta_2 -\Delta_1)(\Delta+\Delta_3-\Delta_4)
\over 2\Delta}
\end{align*}
\begin{align*}
B_2\Big( \Delta_1, \Delta_2, \Delta_3, \Delta_4, \Delta, c \Big) =
{(\Delta+\Delta_2 -\Delta_1 )(\Delta+\Delta_2 -\Delta_1 + 1)
(\Delta+\Delta_3-\Delta_4)(\Delta+\Delta_3-\Delta_4+1)\over 4\Delta(2\Delta+1)} +
\nn\\
+ {\left[(\Delta_1   +\Delta_2 )(2\Delta+1)+\Delta(\Delta-1)
-3(\Delta_1   -\Delta_2 )^2\right]
\left[(\Delta_3+\Delta_4)(2\Delta+1)+\Delta(\Delta-1)
-3(\Delta_3-\Delta_4)^2\right]
\over 2(2\Delta+1)\Big(2\Delta(8\Delta-5) + (2\Delta+1)c\Big)} =
\end{align*}
\begin{align*}
=
{1\over 2\Delta\Big(16\Delta^2+2(c-5)\Delta+c\Big)} \Big[ \
4\Delta(\Delta+2\Delta_2-\Delta_1)(\Delta+2\Delta_3-\Delta_4)(2\Delta+1)+
\nn \\
+
(\Delta+\Delta_2-\Delta_1)(\Delta+\Delta_2-\Delta_1+1)(\Delta+\Delta_3-\Delta_4)
(\Delta+\Delta_3-\Delta_4+1)(4\Delta+c/2)-
\nn \\
-
6\Delta(\Delta+2\Delta_2-\Delta_1)(\Delta+\Delta_3-\Delta_4)
(\Delta+\Delta_3-\Delta_4+1)
\nn\\
-
6\Delta(\Delta+\Delta_2-\Delta_1)(\Delta+\Delta_2-\Delta_1+1)
(\Delta+2\Delta_3-\Delta_4) \ \Big]
\end{align*}
\begin{align*}
B_3\Big( \Delta_1, \Delta_2, \Delta_3, \Delta_4, \Delta, c \Big) =
{1\over 2\Delta (3 \Delta ^2+c \Delta -7 \Delta +2+c)}\left[
(\Delta +3 \Delta_2  -\Delta_1    )(\Delta ^2+3 \Delta +2)
(\Delta +3 \Delta_3 -\Delta_4 ) -\phantom{1\over 2}\right.\\
-2(\Delta +3 \Delta_2  -\Delta_1    )(\Delta +1)
 (\Delta +2 \Delta_3 -\Delta_4 )(\Delta +\Delta_3 -\Delta_4 +2) +\\
 + (\Delta +3 \Delta_2  -\Delta_1    )(\Delta +\Delta_3 -\Delta_4 )
 (\Delta +\Delta_3 -\Delta_4 +1) (\Delta +\Delta_3 -\Delta_4 +2) -\\
 -2(\Delta +2 \Delta_2  -\Delta_1    ) (\Delta +\Delta_2  -\Delta_1    +2)
 (\Delta +1)(\Delta +3 \Delta_3 -\Delta_4 ) + \\
 +(\Delta +\Delta_2  -\Delta_1    ) (\Delta +\Delta_2  -\Delta_1    +1)
 (\Delta +\Delta_2  -\Delta_1    +2)(\Delta +3 \Delta_3 -\Delta_4 )+
\end{align*}
{\fontsize{9pt}{0pt}
$$\vspace{0.3cm}+2\frac{(\Delta +2 \Delta_2  -\Delta_1    )
(\Delta +\Delta_2  -\Delta_1    +2)
 (6 \Delta ^3+9 \Delta ^2-9 \Delta +2 c \Delta ^2+3 c \Delta +c)
 (\Delta +2 \Delta_3 -\Delta_4 )(\Delta +\Delta_3 -\Delta_4 +2)}
 {16\Delta^2+2(c-5)\Delta+c}
 $$ $$ \vspace{0.3cm}-\frac{(\Delta +2 \Delta_2  -\Delta_1    )
 (\Delta +\Delta_2  -\Delta_1    +2)
 (9 \Delta ^2-7 \Delta +3 c \Delta +c)(\Delta +\Delta_3 -\Delta_4 )
 (\Delta +\Delta_3 -\Delta_4 +1)(\Delta +\Delta_3 -\Delta_4 +2)}
 {16\Delta^2+2(c-5)\Delta+c}
 $$ $$\vspace{0.3cm}-\frac{(\Delta +\Delta_2  -\Delta_1    )
 (\Delta +\Delta_2  -\Delta_1    +1)
 (\Delta +\Delta_2  -\Delta_1    +2)(9 \Delta ^2-7 \Delta +3 c \Delta +c)
 (\Delta +2 \Delta_3 -\Delta_4 )(\Delta +\Delta_3 -\Delta_4 +2)}
 {16\Delta^2+2(c-5)\Delta+c}
 $$ $$ \vspace{0.3cm}+ (\Delta +\Delta_2  -\Delta_1    )
 (\Delta +\Delta_2  -\Delta_1    +1)
  (\Delta +\Delta_2  -\Delta_1    +2)
  (\Delta +\Delta_3 -\Delta_4 )(\Delta +\Delta_3 -\Delta_4 +1)
  (\Delta +\Delta_3 -\Delta_4 +2)\times
  $$} \vspace{-0.3cm} $$
  \left.
  \times \frac{
  (24 \Delta ^2-26 \Delta +11 c \Delta +8 c+c^2)
  }
  {12\Big(16\Delta^2+2(c-5)\Delta+c\Big)}\right]
$$
\smallskip\\
The next section is devoted to detailed checks of the equality of coefficients $J_k$ and $B_k$.
Here we briefly comment about the equality of normalization constants $C_{DF}$ and
$C \big( \Delta_1, \Delta_2, \Delta \big) C\big( \Delta, \Delta_3, \Delta_4 \big)$.
Let us make a change (rescaling) of the integration variables:
$z_i = q u_{i}$ for $i \leq N_1$ and $z_i = v_{i}$ for $i > N_1$.
After this rescaling, the Dotsenko-Fateev integral takes the form
\begin{align}
\nonumber Z_{DF} \ = \ &
q^{\dfrac{\alpha_1 \alpha_2}{2 \beta}} \ (1 - q)^{\dfrac{\alpha_2 \alpha_3}{2 \beta}}
\int\limits_{0}^{1} du_1 \ldots \int\limits_{0}^{1} du_{N_1}
\ \int\limits_{0}^{1} dv_1 \ldots \int\limits_{0}^{1} dv_{N_2}
\prod\limits_{i < j} (q u_j - q u_i)^{2\beta}
\prod\limits_{i < j} (v_j - v_i)^{2\beta} \times \\
&\prod\limits_{i = 1}^{N_1} \prod\limits_{j = 1}^{N_2} (v_j - q u_i)^{2\beta}
\prod\limits_{i = 1}^{N_1} (q u_i)^{\alpha_1} (q u_i - q)^{\alpha_2} (q u_i - 1)^{\alpha_3}
\prod\limits_{j = 1}^{N_2} v_i^{\alpha_1} (v_i - q)^{\alpha_2} (v_i - 1)^{\alpha_3}
\end{align}
\smallskip\\
Since the integration limits no longer depend on $q$,
it is now easy to find the overall $q$-degree of the integral
\begin{align}
{\rm deg}_{DF} \ = \ \beta N_1(N_1-1) + N_1 + \alpha_1 N_1 + \alpha_2 N_1
+ \dfrac{\alpha_1 \alpha_2}{2 \beta} \
\mathop{=}^{(\ref{Relations1})} \ \Delta - \Delta_1 - \Delta_2
\end{align}
in accordance with (\ref{MainEquality}).

\bigskip

It is also easy to find the normalization constant:
\be
C_{DF} = C_{N_1}(\alpha_1,\alpha_2) C_{N_2}(a,\alpha_3)
\ee
where $a = \alpha_1 + \alpha_2 + 2 \beta N_1$ and
\be
C_{N}(x,y) = \prod\limits_{i = 1}^{N} \int\limits_{0}^{1} du_i
\prod\limits_{i < j} (u_j - u_i)^{2\beta}
\prod\limits_{i = 1}^{N} u_i^{x} (u_i - 1)^{y} =
\nn \\ =
\prod\limits_{k = 1}^{N} \dfrac{\Gamma(x + 1 + \beta(k-1))
\Gamma(y + 1 + \beta(k-1))\Gamma(1 + \beta k)}
{\Gamma(x + y + 2 + (N+k-2)\beta)\Gamma(\beta + 1)}
\ee
is the so-called Selberg integral \cite{DVagt,Selberg,Wilma}.
As one can see, $C_{DF}$ is a product of two factors,
which are associated with the two vertices of the diagram
and depend on respective screening multiplicities ($N_1$ and $N_2$)
and incoming dimensions ($\alpha_1,\alpha_2$ and $\alpha,\alpha_3$)
of these vertices.
As demonstrated in  s.4.1 of ref.\cite{Wilma}, $C_{N}(x,y)$ coincides
with the perturbative Nekrasov function and is similar
(though not literally equal to, see s.\ref{prob} below)
to the three-point function on a sphere $C(\Delta_1,\Delta_2,\Delta_3)$
also known as DOZZ three-point function \cite{DOZZ}.
Therefore, we conclude that the Dotsenko-Fateev integral reproduces
the conformal block modulo a constant factor.
Hence we do not consider this factor in what follows,
and concentrate on the coefficients of the series expansions (\ref{Expansion1})
and (\ref{Expansion2}).

\section{Evidence in support of (8)}

\subsection{The case of $\beta = 1$ and all $\alpha_i$ vanishing}

This is the simplest possible situation,
where the equivalence between the Dotsenko-Fateev partition function
and the 4-point conformal block can be seen.
For the sake of brevity, we denote $X = N_2^2 + 2 N_1 N_2$.
Direct calculation gives:

\be
J_1(0,0,0,N_1,N_2,\beta = 1) = - \dfrac{X}{2}
\ee

\be
J_2(0,0,0,N_1,N_2,\beta = 1) = - \dfrac{1 - 7 N_1+ 12 N_1^2}{4(4N_1^2 - 1)^2} X
+ \dfrac{1 - 3N_1^2 + 8N_1^4}{4(4N_1^2 - 1)^2} X^2
\ee

\be
J_3(0,0,0,N_1,N_2,\beta = 1) =  - \dfrac{4 - 26 N_1 + 40 N_1^2}{24(4N_1^2 - 1)^2} X
+ \dfrac{6 - 15 N_1 + 36 N_1^2}{{24(4N_1^2 - 1)^2}} X^2
- \dfrac{2 - N_1 + 8 N_1^2}{{24(4N_1^2 - 1)^2}} X^3
\ee
\smallskip\\
The integral at level 4 is also easily calculated (the corresponding conformal block
is directly obtained too):

\begin{align}
\nonumber J_4(0,0,0,N_1,N_2,\beta = 1) = &
- \dfrac{ 3888-27540 N_1^2+56292 N_1^4-35040 N_1^6+6720 N_1^8 }
{384(4N_1^2 - 1)^2(4N_1^2 - 9)^2} X + \\ \nonumber  & \\
\nonumber & + \dfrac{ 7101-22305 N_1^2+50180 N_1^4-33904 N_1^6+6848 N_1^8 }
{384(4N_1^2 - 1)^2(4N_1^2 - 9)^2} X^2 - \\ \nonumber  & \\
\nonumber & - \dfrac{ 3834-4794 N_1^2+13632 N_1^4-10656 N_1^6+2304 N_1^8 }
{384(4N_1^2 - 1)^2(4N_1^2 - 9)^2} X^3 + \\ \nonumber  & \\
& +  \dfrac{ 621-201 N_1^2+1132 N_1^4-1088 N_1^6+256 N_1^8 }
{384(4N_1^2 - 1)^2(4N_1^2 - 9)^2} X^4
\end{align}
\smallskip\\
Comparing these expressions with the conformal block with vanishing
$\Delta_1, \Delta_2$ and $\Delta_3$, one finds

\begin{align}
J_k(0,0,0,N_1,N_2,\beta = 1) =
B_k\Big(0,0,0, (N_1 + N_2)^2, N_1^2 , c = 1\Big),
\label{rela1}
\end{align}
\smallskip\\
at least for $k = 1,2,3$. Thus, in this particular case the Dotsenko-Fateev partition function
correctly reproduces the conformal block up to level 3, at least.
This is the first evidence in support of (\ref{MainEquality}).
This particular relation (\ref{rela1}) has been already verified
up to level 2 in \cite{MMSh1}.

\subsection{The case of $\beta = 1$ and non-vanishing $\alpha_1$}

As $\alpha_i$ are switched on, the computation of coefficients $J_k$ becomes harder.
Currently, the two levels are calculated:

\begin{align}
J_1(\alpha_1,0,0,N_1,N_2,\beta = 1) = \dfrac{-2 N_{1} N_{2} (N_{1}+\alpha_{1}) (2 N_{1}+N_{2}+\alpha_{1})}{(2 N_{1}+\alpha_{1})^2}
\end{align}

\begin{align}
\nonumber J_2(\alpha_1,0,0,N_1,N_2,\beta = 1) \ = \ & \dfrac{N_{1}N_{2}(N_{1}+\alpha_{1})(2 N_{1}+N_{2}+\alpha_{1})}{(\alpha_{1}+1+2 N_{1})^2(2 N_{1}+\alpha_{1})^2(\alpha_{1}-1+2 N_{1})^2} \times \Big( -1+2 \alpha_{1}^2+N_{2} \alpha_{1}+N_{2}^2+\\ \nonumber & \\ \nonumber & + 7 N_{1} \alpha_{1}+2 N_{1} N_{2}+7 N_{1}^2-\alpha_{1}^4-N_{2} \alpha_{1}^3-N_{2}^2 \alpha_{1}^2-7 N_{1} \alpha_{1}^3-5 N_{1} N_{2} \alpha_{1}^2-\\ \nonumber & \\ \nonumber &-3 N_{1} N_{2}^2 \alpha_{1}-19 N_{1}^2 \alpha_{1}^2-9 N_{1}^2 N_{2} \alpha_{1}-3 N_{1}^2 N_{2}^2-24 N_{1}^3 \alpha_{1}-6 N_{1}^3 N_{2}-\\ \nonumber & \\ \nonumber &-12 N_{1}^4+2 N_{1} N_{2} \alpha_{1}^4+2 N_{1} N_{2}^2 \alpha_{1}^3+14 N_{1}^2 N_{2} \alpha_{1}^3+10 N_{1}^2 N_{2}^2 \alpha_{1}^2+\\ \nonumber & \\ &+36 N_{1}^3 N_{2} \alpha_{1}^2+16 N_{1}^3 N_{2}^2 \alpha_{1}+40 N_{1}^4 N_{2} \alpha_{1}+8 N_{1}^4 N_{2}^2+16 N_{1}^5 N_{2} \Big)
\end{align}
\smallskip\\
Comparing these expressions with the conformal block with vanishing
$\Delta_2, \Delta_3$ (and $\Delta_1$ non-vanishing) one finds

\begin{align}
J_k(\alpha_1,0,0,N_1,N_2,\beta = 1) =
B_k\left( \dfrac{\alpha_1^2}{4}, 0, 0, \left(N_1 + N_2
+ \dfrac{\alpha_1}{2} \right)^2, \left(N_1 + \dfrac{\alpha_1}{2}\right)^2 ,
c = 1 \right), \ \ \ \ k = 1,2
\end{align}
\smallskip\\
One can see that, in this particular case, the
Dotsenko-Fateev partition function reproduces correctly the conformal
block up to level 2.
This is yet another evidence in support of (\ref{MainEquality}).

\subsection{The case of $\beta = 1$ and non-vanishing $\alpha_2$ or $\alpha_3$}

Similarly, for non-vanishing $\alpha_2$ one finds

\begin{align}
J_1(0,\alpha_2,0,N_1,N_2,\beta = 1) =
-\dfrac{N_2(2 N_{1}^2+2 N_{1} \alpha_{2}+\alpha_{2}^2) (2 N_{1}+N_{2}+\alpha_{2})}{(2 N_{1}+\alpha_{2})^2}
\end{align}

\begin{align}
\nonumber & J_2(0,\alpha_2,0,N_1,N_2,\beta = 1) \ = \
\dfrac{N_{2}(2 N_{1}+N_{2}+\alpha_{2})}{2(1+\alpha_{2}+2 N_{1})^2(2 N_{1}+\alpha_{2})^2(-1+\alpha_{2}+2 N_{1})^2} \times \Big( -\alpha_{2}^2-2 N_{1} \alpha_{2}-2 N_{1}^2+ \\ \nonumber & \\ \nonumber & +2 \alpha_{2}^4+N_{2} \alpha_{2}^3+N_{2}^2 \alpha_{2}^2+12 N_{1} \alpha_{2}^3+4 N_{1} N_{2} \alpha_{2}^2+2 N_{1} N_{2}^2 \alpha_{2}+26 N_{1}^2 \alpha_{2}^2 + 6 N_{1}^2 N_{2} \alpha_{2}+2 N_{1}^2 N_{2}^2+28 N_{1}^3 \alpha_{2}+\\ \nonumber & \\ \nonumber &+ 4 N_{1}^3 N_{2}+14 N_{1}^4 - \alpha_{2}^6-2 N_{2} \alpha_{2}^5  - 2 N_{2}^2 \alpha_{2}^4-10 N_{1} \alpha_{2}^5-14 N_{1} N_{2} \alpha_{2}^4-10 N_{1} N_{2}^2 \alpha_{2}^3-40 N_{1}^2 \alpha_{2}^4 - 36 N_{1}^2 N_{2} \alpha_{2}^3 - \\ \nonumber & \\ \nonumber &- 16 N_{1}^2 N_{2}^2 \alpha_{2}^2-84 N_{1}^3 \alpha_{2}^3-44 N_{1}^3 N_{2} \alpha_{2}^2 - 12 N_{1}^3 N_{2}^2 \alpha_{2} - 102 N_{1}^4 \alpha_{2}^2-30 N_{1}^4 N_{2} \alpha_{2} - 6 N_{1}^4 N_{2}^2-72 N_{1}^5 \alpha_{2}-12 N_{1}^5 N_{2} \\ \nonumber & \\ \nonumber & -24 N_{1}^6+N_{2} \alpha_{2}^7+N_{2}^2 \alpha_{2}^6+10 N_{1} N_{2} \alpha_{2}^6 + 8 N_{1} N_{2}^2 \alpha_{2}^5+44 N_{1}^2 N_{2} \alpha_{2}^5+28 N_{1}^2 N_{2}^2 \alpha_{2}^4+112 N_{1}^3 N_{2} \alpha_{2}^4+56 N_{1}^3 N_{2}^2 \alpha_{2}^3+\\ \nonumber & \\ & + 180 N_{1}^4 N_{2} \alpha_{2}^3+68 N_{1}^4 N_{2}^2 \alpha_{2}^2+184 N_{1}^5 N_{2} \alpha_{2}^2 + 48 N_{1}^5 N_{2}^2 \alpha_{2}+112 N_{1}^6 N_{2} \alpha_{2}+16 N_{1}^6 N_{2}^2+32 N_{1}^7 N_{2} \Big)
\end{align}
\smallskip\\
Comparing these expressions with the conformal block with vanishing
$\Delta_1, \Delta_3$ (and $\Delta_2$ non-vanishing) one finds

\begin{align}
J_k(0,\alpha_2,0,N_1,N_2,\beta = 1) = B_k\left( 0, \dfrac{\alpha_2^2}{4}, 0,
\left(N_1 + N_2 + \dfrac{\alpha_2}{2} \right)^2, \left(N_1
+ \dfrac{\alpha_2}{2}\right)^2 , c = 1 \right), \ \ \ \ k = 1,2
\label{Rel1}
\end{align}
\smallskip\\
For non-vanishing $\alpha_3$ one finds

\begin{align}
J_1(0,0,\alpha_3,N_1,N_2,\beta = 1) = -\dfrac{1}{2} N_{2} \alpha_{3}-\dfrac{1}{2} N_{2}^2-\dfrac{1}{2} N_{1} \alpha_{3}-N_{1} N_{2}
\end{align}

\begin{align}
\nonumber & J_2(0,0,\alpha_3,N_1,N_2,\beta = 1) \ = \ \dfrac{1}{4(1+2 N_1)^2 (1-2 N_1)^2}
\Big( -N_{2} \alpha_{3}-N_{2}^2-N_{1} \alpha_{3}-2 N_{1} N_{2}+N_{2}^2 \alpha_{3}^2+2 N_{2}^3 \alpha_{3}+\\ \nonumber & \\ \nonumber &+ N_{2}^4+2 N_{1} N_{2} \alpha_{3}^2+6 N_{1} N_{2}^2 \alpha_{3}+4 N_{1} N_{2}^3+11 N_{1}^2 N_{2} \alpha_{3}+11 N_{1}^2 N_{2}^2+7 N_{1}^3 \alpha_{3}+14 N_{1}^3 N_{2}-3 N_{1}^2 N_{2}^2 \alpha_{3}^2-\\ \nonumber & \\ \nonumber &-6 N_{1}^2 N_{2}^3 \alpha_{3}-3 N_{1}^2 N_{2}^4-6 N_{1}^3 N_{2} \alpha_{3}^2-18 N_{1}^3 N_{2}^2 \alpha_{3}-12 N_{1}^3 N_{2}^3-2 N_{1}^4 \alpha_{3}^2-24 N_{1}^4 N_{2} \alpha_{3}-24 N_{1}^4 N_{2}^2-\\ \nonumber & \\ \nonumber &-12 N_{1}^5 \alpha_{3}-24 N_{1}^5 N_{2}+8 N_{1}^4 N_{2}^2 \alpha_{3}^2+16 N_{1}^4 N_{2}^3 \alpha_{3}+8 N_{1}^4 N_{2}^4+16 N_{1}^5 N_{2} \alpha_{3}^2+48 N_{1}^5 N_{2}^2 \alpha_{3}+\\ \nonumber & \\ \nonumber &+32 N_{1}^5 N_{2}^3+8 N_{1}^6 \alpha_{3}^2+32 N_{1}^6 N_{2} \alpha_{3}+32 N_{1}^6 N_{2}^2 \Big)
\end{align}
Comparing these expressions with the conformal block with vanishing
$\Delta_1, \Delta_2$ (and $\Delta_3$ non-vanishing) one finds

\begin{align}
J_k(0,0,\alpha_3,N_1,N_2,\beta = 1) = B_k\left( 0, 0, \dfrac{\alpha_3^2}{4},
\left(N_1 + N_2 + \dfrac{\alpha_3}{2} \right)^2, N_1^2 , c = 1 \right),
\ \ \ \ k = 1,2
\label{Rel2}
\end{align}
\smallskip\\
Relations (\ref{Rel1}) and (\ref{Rel2}) provide yet another evidence in support of
(\ref{MainEquality}).

\subsection{The case of $\beta = 1$ and all $\alpha_i$ non-vanishing }

For non-vanishing $\alpha_1, \alpha_2$ and $\alpha_3$ one finds
\begin{align}
\nonumber J_1(\alpha_1,\alpha_2,\alpha_3,N_1,N_2,\beta = 1) \ = \
& - \dfrac{(2 N_{1}^2+2 N_{1} \alpha_{1}+2 N_{1} \alpha_{2}+\alpha_{2}
\alpha_{1}+\alpha_{2}^2)}{2(2 N_{1}+\alpha_{1}+\alpha_{2})^2}
\ \times \emph{} \\
\nonumber & \\ \times & \Big( 4 N_{1} N_{2}+2 N_{1}
\alpha_{3}+2 N_{2}^2+2 N_{2} \alpha_{3}+\alpha_{3} \alpha_{1}+2 N_{2}
\alpha_{1}+\alpha_{3} \alpha_{2}+2 N_{2} \alpha_{2} \Big)
\end{align}
The formula for $J_2(\alpha_1,\alpha_2,\alpha_3,N_1,N_2,\beta = 1)$
is a little lengthy, even for $\beta=1$.
For the sake of completeness it is presented in the separate Appendix
at the end of this paper.
Comparing these expressions with the conformal block, one finds
\begin{align}
\!\!\! J_k\Big(\alpha_1,\alpha_2,\alpha_3,N_1,N_2, \beta = 1\Big) =
 B_k\!\left( \dfrac{\alpha_1^2}{4}, \dfrac{\alpha_2^2}{4},
 \dfrac{\alpha_3^2}{4}, \left(N_1 + N_2
 + \dfrac{\alpha_1 + \alpha_2 + \alpha_3}{2} \right)^2\!\!\!, \left(N_1
 + \dfrac{\alpha_1 + \alpha_2}{2}\right)^2\!\!\!,\, c = 1 \right)
\end{align}
at least, for $k = 1,2$.
This relation is quite important:
the equivalence between the conformal block and the Dotsenko-Fateev integral
continues to hold, when the external dimensions are non-vanishing and arbitrary.
Given this relation, one obtains the following correspondence between the parameters:
\be
\left\{
\begin{array}{lllll}
\Delta_1 = \dfrac{\alpha_1^2}{4}, \ \ \ \Delta_2 = \dfrac{\alpha_2^2}{4}, \ \ \
\Delta_3 = \dfrac{\alpha_3^2}{4} \\
\\
\Delta_4 = \left(N_1 + N_2 + \dfrac{\alpha_1 + \alpha_2 + \alpha_3}{2} \right)^2 \\
\\
\Delta = \left(N_2 + \dfrac{\alpha_1 + \alpha_2}{2} \right)^2\ \
\\
\\
(\beta = 1) \ \leftrightarrow \ (c = 1) \\
\end{array}
\right.
\ee
Of course, in this particular form it is valid only for $\beta = 1$;
we now proceed to the generalization to $\beta \neq 1$.

\subsection{The case of arbitrary $\beta$ and all $\alpha_i$ non-vanishing }

To generalize the above formulas to an arbitrary $\beta$,
it is instructive first to look at the level 1 coefficient $J_1(\beta)$
with all $\alpha_i$ non-vanishing.
The corresponding expression is still not very complicated and,
most importantly, factorized:
\begin{align}
& \nonumber J_1(\alpha_1,\alpha_2,\alpha_3,N_1,N_2,\beta) \ =
\ - \dfrac{2 \beta N_{1}-2 \beta^2 N_{1}+2 N_{1}^2 \beta^2+2 N_{1}
\beta \alpha_{1}+2 \alpha_{2}-2 \beta \alpha_{2}+2 N_{1}
\beta \alpha_{2}+\alpha_{2} \alpha_{1}+\alpha_{2}^2}{2
\beta (2 \beta N_{1}+\alpha_{1}+\alpha_{2}) (2 \beta N_{1}+\alpha_{1}+\alpha_{2}
-2 \beta+2)} \times \emph{} \\
& \Big( 4 N_{1} N_{2} \beta^2+2 N_{1} \beta \alpha_{3}
-2 N_{2} \beta^2+2 \beta^2 N_{2}^2+2 \beta N_{2} \alpha_{1}
+2 N_{2} \beta \alpha_{3}+2 \beta N_{2}+2 N_{2} \beta \alpha_{2}
+\alpha_{3} \alpha_{1}+\alpha_{3} \alpha_{2} \Big)
\end{align}
It is easy to recognize here the level 1 conformal block:
\begin{align}
 J_1(\alpha_1,\alpha_2,\alpha_3,N_1,N_2, \beta) =
 B_1(\Delta_1,\Delta_2,\Delta_3,\Delta_4,\Delta,c)
\end{align}
with the following relation between parameters:
\begin{align}
\left\{
\begin{array}{lllll}
\Delta_1 = \dfrac{ \alpha_1 ( \alpha_1 + 2 - 2 \beta )}{4 \beta} \\
 \\
 \Delta_2 = \dfrac{ \alpha_2 ( \alpha_2 + 2 - 2 \beta )}{4 \beta} \\
  \\
  \Delta_3 = \dfrac{ \alpha_3 ( \alpha_3 + 2 - 2 \beta )}{4 \beta} \\
\\
\Delta_4 = \dfrac{(2 \beta (N_1 + N_2 ) + \alpha_1 + \alpha_2 + \alpha_3 )
(2 \beta (N_1 + N_2) + \alpha_1 + \alpha_2 + \alpha_3 + 2-2 \beta)}{4\beta} \\
\\
\Delta = \dfrac{(2 \beta N_2 + \alpha_1 + \alpha_2 )
(2 \beta N_2 + \alpha_1 + \alpha_2 + 2 - 2 \beta )}{4 \beta}
\\
\end{array}
\right.
\end{align}
As one can see, up to level 1 the Dotsenko-Fateev partition function coincides
with the conformal block even for $\beta \neq 1$ (or, what is the same, for $c \neq 1$).
However, the precise correspondence between $\beta$ and $c$ is not seen from the above
relations. We establish this correspondence below.

\subsection{ Determination of the central charge }
The central charge $c$ is not constrained by any of the relations above:
it can not be found from level 1 considerations, since $B_1$ does not depend on $c$.
To find $c$, one needs to look at
$$ J_2(0,0,0,N_1,N_2, \beta) =
\dfrac{N_2(1+\beta N_2+2 \beta N_1-\beta)}
{4(3+2 \beta N_1-2 \beta)(2 N_1+1)(2+2 \beta N_1-3 \beta)(-1+2 \beta N_1)} \times
$$
\begin{center}
$\times \Big(6-12 N_1^4 \beta^3+19 \beta^2 N_1^2+11 \beta^2 N_1-6 N_1 \beta^3
-7 \beta^3 N_2^2 N_1^2-21 N_1^2 N_2 \beta^3+16 N_2 N_1^5 \beta^4
+8 N_2^2 N_1^4 \beta^4+6 \beta^3 N_2-4 N_1 N_2 \beta^2+9 N_1 N_2 \beta^3
+9 N_1 N_2 \beta-46 \beta^3 N_2 N_1^3+6 N_1-6 N_2+15 N_1 N_2^2 \beta^2
-15 N_1 N_2^2 \beta^3+21 N_1^2 N_2 \beta^2+2 \beta^4 N_2^2 N_1^2-13 \beta
+6 \beta^2-11 \beta N_1+40 N_2 N_1^4 \beta^3+20 \beta^4 N_2 N_1^3-6 N_1 N_2 \beta^4
+10 N_1^2 N_2 \beta^4+6 N_1 N_2^2 \beta^4-6 N_1 N_2-6 N_1^2 \beta^3+24 \beta^3 N_1^3
-6 N_1^2 \beta-24 N_1^3 \beta^2-6 N_2^2 \beta^3-19 N_2 \beta^2-10 N_1^2 N_2 \beta
+20 N_1^3 N_2 \beta^2-40 N_1^4 N_2 \beta^4-6 N_2^2 \beta-6 N_1 N_2^2 \beta
+2 N_1^2 N_2^2 \beta^2-16 N_1^3 N_2^2 \beta^4+16 N_1^3 N_2^2 \beta^3+19 \beta N_2
+13 \beta^2 N_2^2\Big)
$\smallskip\\\end{center}
This has to be compared with
\begin{align}
B_2(0,0,0,\Delta_4,\Delta, c) = \dfrac{(\Delta-\Delta_4)
(8 \Delta^3+\Delta^2 c+8 \Delta^2-8 \Delta^2 \Delta_4+
2 c \Delta-\Delta_4 \Delta c+4 \Delta_4 \Delta-8 \Delta+c-\Delta_4 c)}
{4 (16 \Delta^2-10 \Delta+2 c \Delta+c)}
\end{align}
where we put
\begin{align}
\Delta_4 = \beta (N_1 + N_2) \left( N_1 + N_2 + \beta + \dfrac{1}{\beta} \right),
\ \ \ \ \Delta = \beta N_2 \left( N_2 + \beta + \dfrac{1}{\beta} \right)
\end{align}
After that it is easy to see that the difference $J_1 - B_1$
is divisible by $6 \beta^2+ \beta c-13 \beta+6$. Therefore,
\be
c(\beta) = 13 - 6 \beta - \dfrac{6}{\beta} = 1 - 6 \left( \sqrt{\beta}
- \dfrac{1}{\sqrt{\beta}} \right)^2
\ee
This completes our check of the relation (\ref{MainEquality}).
Of course, much more evidence can be gathered: say,
one can calculate the whole $J_2(\alpha_1, \alpha_2, \alpha_3,N_1,N_2, \beta) $
and compare it with $B_2(\Delta_1,\Delta_2,\Delta_3,\Delta_4,\Delta, c)$,
and then continue to level 3 and higher.
However, the summary of evidence above available at the moment
is already quite convincing.
Instead of doing further more involved computer calculations,
it is desirable to find a clever theoretical proof of (\ref{MainEquality}).

\subsection{The lesson}

The main lesson so far is that, at least, the $4$-point spherical conformal block
{\it for arbitrary values of the five conformal dimensions} is given by the
free field correlator (\ref{FD4pt}), with two new parameters $N_1$ and $N_2$,
which are used to lift the two restrictions of the naive free field:
on the intermediate $\alpha$-parameter and on the sum of the four external
$\alpha$-parameters. Introduction of $N_1$ and $N_2$ extends the
$3$-dimensional space of the naive free field conformal $4$-point blocks
and converts it into the $5$-dimensional space of arbitrary $4$-point
conformal blocks. The central charge is of course arbitrary.
In order to describe the entire moduli space of conformal blocks, the
parameters $N_1$ and $N_2$ should be arbitrary, not obligatory positive integers,
thus, the procedure includes an analytical continuation in $N_1$ and $N_2$.

As we shall see in s.\ref{multi},
this result can be straightforwardly extended to spherical conformal
blocks with arbitrary number of legs.
Extension to higher genera requires
more work, only a preliminary result for 1-point function on a torus
will be presented in s.\ref{tor} below.
Extensions in two other directions: from the Virasoro to $W$ chiral algebras
(from $U(2)$ quivers to $U(N)$ quivers)
and from generic to degenerate Verma modules
should be straightforward, but are not considered in the present paper.

\subsection{A problem \label{prob}}

\begin{figure}\begin{center}
\includegraphics[totalheight=130pt]{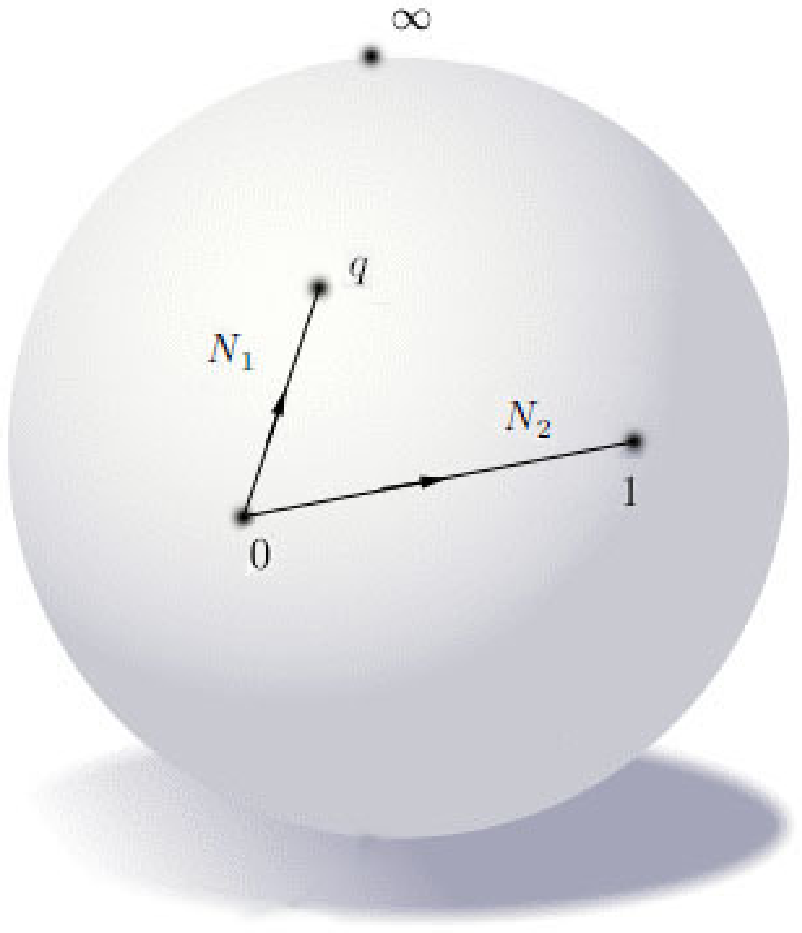}
\includegraphics[totalheight=130pt]{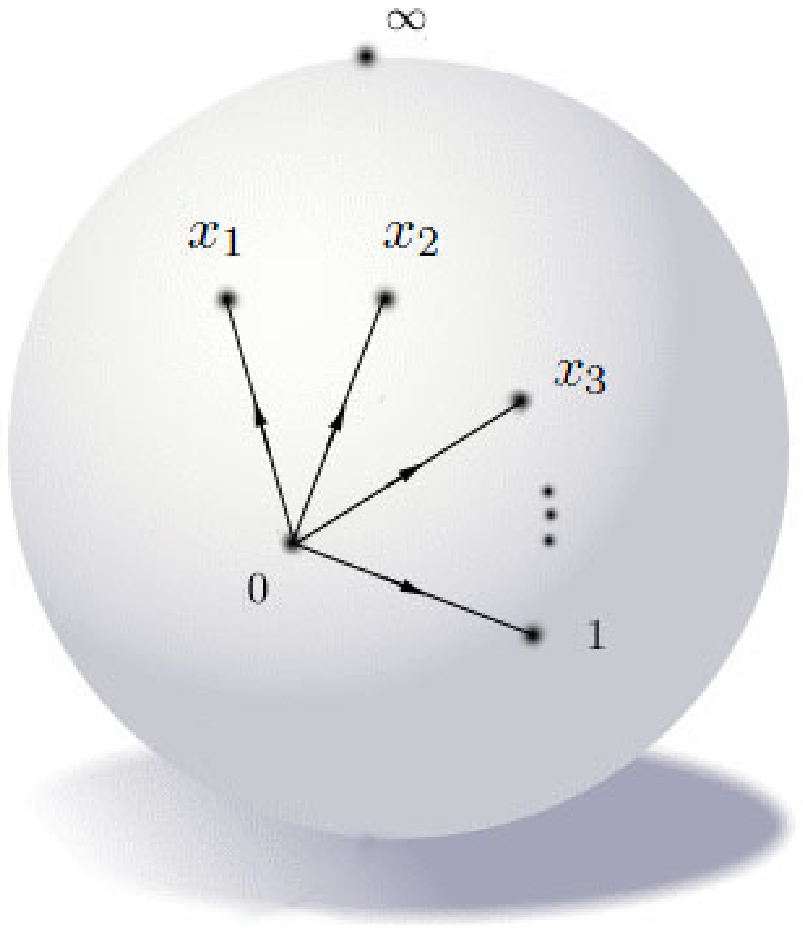}
\caption{\footnotesize
}
\label{spheric45}
\end{center}\end{figure}

\begin{figure}\begin{center}
\includegraphics[totalheight=130pt]{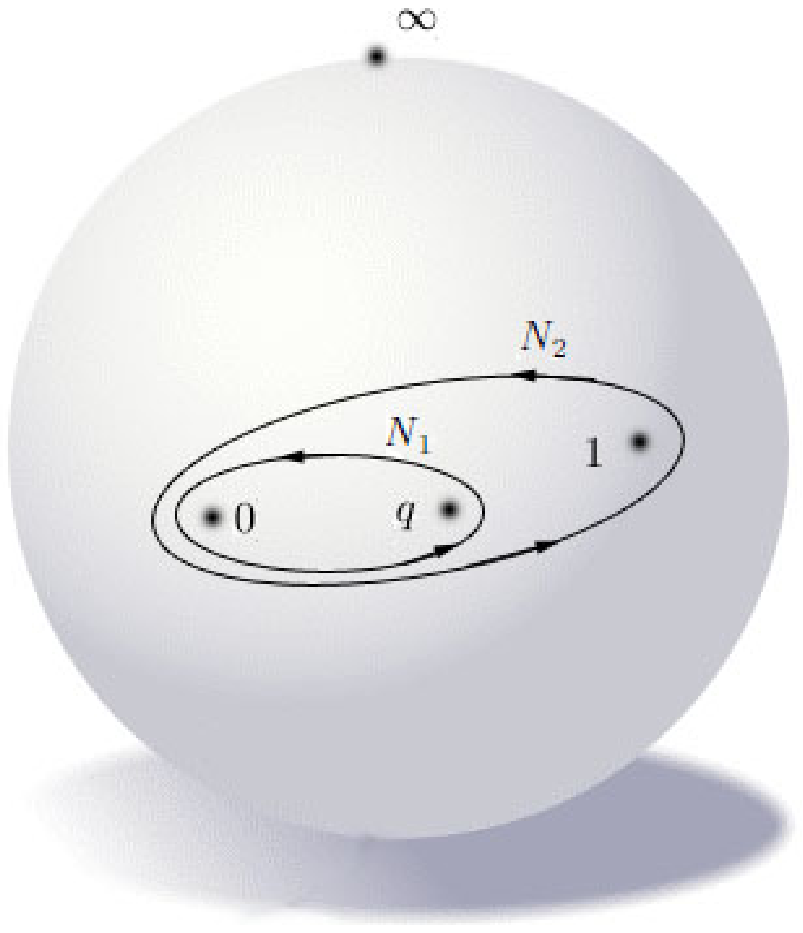}
\includegraphics[totalheight=130pt]{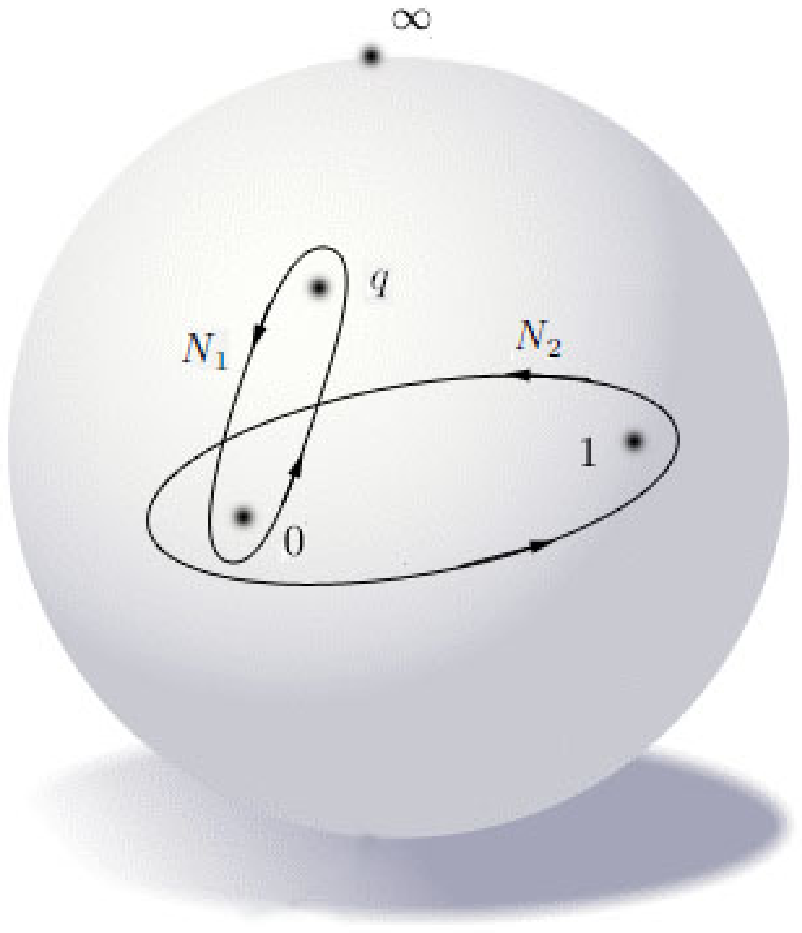}
\caption{\footnotesize
}
\label{spheric12}
\end{center}\end{figure}

\begin{figure}\begin{center}
\includegraphics[totalheight=120pt]{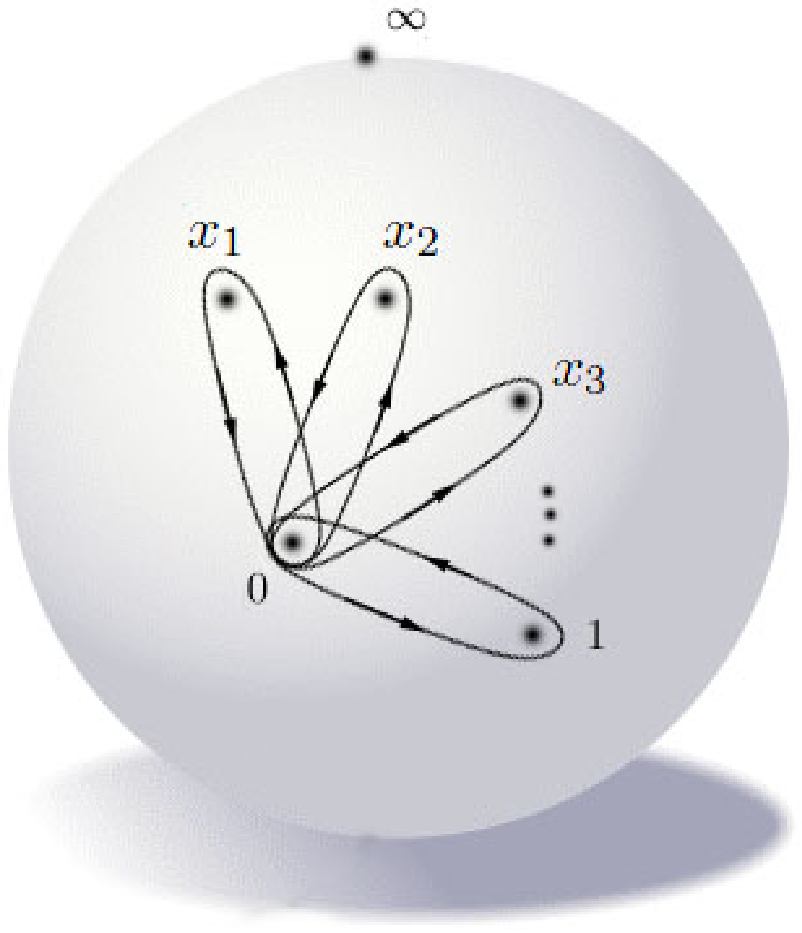}
\caption{\footnotesize
}
\label{spheric3}
\end{center}
\end{figure}

\begin{figure}
\begin{center}
\includegraphics[totalheight=110pt]{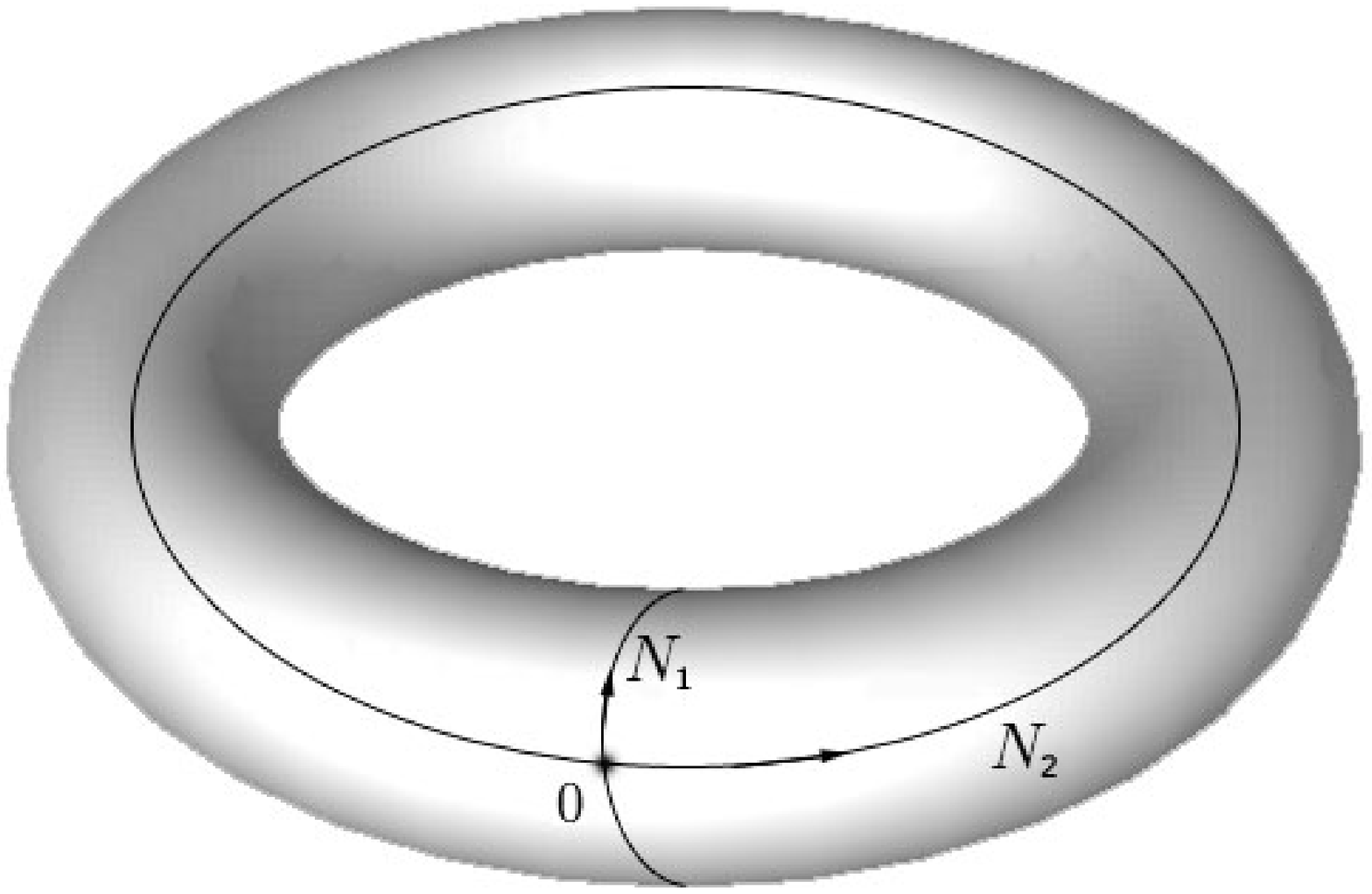}
\caption{\footnotesize
}
\label{toric}
\end{center}\end{figure}

Despite undisputable success, there remains a problem, which still needs
to be resolved: it concerns the issue of integration contours.
The thing is  that in our approach we integrate {\it polynomials}
and then perform an analytical continuation to arbitrary values of parameters
$\tilde\alpha, b, N$.
A polynomial can be integrated only along {\it segments},
Fig.\ref{spheric45}:
while integrals along the closed contours vanish because of the factors like
$1-e^{2\pi i\alpha}$ in
\be
\int_{{{\rm around\ a\ cut}}\atop{{\rm between\ 0\ and\ 1}}}
[z(1-z)]^\alpha =
\left(1-e^{2\pi i\alpha}\right)\int_0^1[z(1-z)]^\alpha
\ee
Of course, after the analytical continuation the answer should be
represented by a closed-contour integral, but our technique can not distinguish,
for example, between the two options in Fig.\ref{spheric12}.
The choice in favor of the right picture
is obvious in the case of
the 4-point function, but the (correct) choice of Fig.\ref{spheric3}
for analytical continuation of the right picture in Fig.\ref{spheric45}
is somewhat less motivated.
Even harder problems arise in the case of a torus, Fig.\ref{toric},
because there we need to integrate a holomorphic Green function,
which is not periodic along the $B$-cycle.

This lack of understanding does not allow us to provide a complete description
of the structure constants $C$ at the end of s.2
(note that in \cite{Wilma} also only a segment integral is considered,
and this contributes to a difference between $C_N(\alpha_1,\alpha_2)$
and the DOZZ function).

\begin{figure}\begin{center}
\includegraphics[totalheight=170pt]{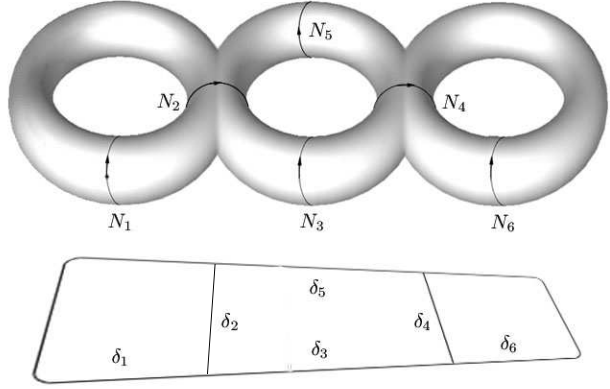}
\caption{\footnotesize
An example of a genus three Riemann surface with
$3g-3=6$ non-{\it homotopic} non-contractable contours
and the associated Feynman-like vacuum diagram for the conformal
block with $3g-3=6$ internal dimensions and no external legs.}
\label{highgen}
\end{center}\end{figure}

Even more important, this obscures generalization to higher genera:
only the case of $N_2=0$ for a torus will be briefly considered in s.5.
An additional problem for genus $g>1$ is a mismatch between the number $3g-3$
of internal dimensions in the vacuum diagram and the number $2g$ of non-{\it homological}
cycles. It can happen that the Dotsenko-Fateev integrals are taken along all the
$3g-3$ non-{\it homotopic} cycles, see Fig.\ref{highgen},
but again the issue of analytical continuation in the language of contours
should be better understood to justify this hypothesis.
An additional difficulty here is the lack of knowledge about conformal blocks
beyond genus one.

These problems are essential and their resolution is one of the primary
tasks in further development along the lines of the present paper.

\section{Multipoint functions on a sphere \label{multi}}

\begin{figure}\begin{center}
\unitlength 1mm 
\linethickness{0.4pt}
\ifx\plotpoint\undefined\newsavebox{\plotpoint}\fi 
\begin{picture}(100,30)(0,-5)
%
%
\put(0,10){\line(1,0){45}}
\put(55,10){\line(1,0){45}}
\put(10,10){\line(0,1){10}}
\put(30,10){\line(0,1){10}}
\put(70,10){\line(0,1){10}}
\put(90,10){\line(0,1){10}}
\put(50,10){\makebox(0,0)[cc]{$\ldots$}}
\put(-8,10){\makebox(0,0)[cc]{$V_{\Delta_1}(0)$}}
\put(8,25){\makebox(0,0)[cc]{$V_{\Delta_2}(x_{1})$}}
\put(30,25){\makebox(0,0)[cc]{$V_{\Delta_3}(x_{2})$}}
\put(70,25){\makebox(0,0)[cc]{$V_{\Delta_{m-2}}(x_{m-3})$}}
\put(92,25){\makebox(0,0)[cc]{$V_{\Delta_{m-1}}(1)$}}
\put(110,10){\makebox(0,0)[cc]{$V_{\Delta_m}(\infty)$}}
\put(20,15){\makebox(0,0)[cc]{$\delta_{1}$}}
\put(40,15){\makebox(0,0)[cc]{$\delta_{2}$}}
\put(60,15){\makebox(0,0)[cc]{$\delta_{m-4}$}}
\put(80,15){\makebox(0,0)[cc]{$\delta_{m-3}$}}
\put(10,10){\circle{4}}
\put(10,4){\makebox(0,0)[cc]{$N_1$}}
\put(30,10){\circle{4}}
\put(29,4){\makebox(0,0)[cc]{$N_2$}}
\put(70,10){\circle{4}}
\put(74,4){\makebox(0,0)[cc]{$N_{m-3}$}}
\put(90,10){\circle{4}}
\put(92,4){\makebox(0,0)[cc]{$N_{m-2}$}}
\put(12,-1){\makebox(0,0){$\tilde a_1 = \tilde\alpha_1+\tilde\alpha_2 + bN_1$}}
\put(50,1){\makebox(0,0){$\tilde a_2 = \tilde a_1+\tilde\alpha_3+bN_2 =$}}
\put(55,-3){\makebox(0,0){$ = \tilde \alpha_1+\tilde\alpha_2+\tilde\alpha_3+b(N_1+N_2)$}}
\put(18,2){\vector(1,3){2}}
\put(44,4){\vector(-1,3){1.6}}
\end{picture}
\caption{\footnotesize{
Feynman-like diagram for a comb-like conformal block.
Here $\delta_j = \tilde{a}_j\left(\tilde{a}_j + \frac{1}{b}-b\right)$ and $\Delta_j
= \tilde\alpha_j\left(\tilde\alpha_j + \frac{1}{b}-b\right)$ are the $m - 3$
internal and $m$ external dimensions, respectively, and $x_i = \prod_{j=i}^{m-3} q_j$.
The role of the screenings is to modify the free field selection rule
at the vertices: instead of $\tilde{a}_{j} = \tilde{a}_{j-1} + \tilde\alpha_{j+1}$
one has $\tilde{a}_{j} = \tilde{a}_{j-1} + \tilde\alpha_{j+1} + b N_j$.
}}\label{blockCircles}
\end{center}\end{figure}
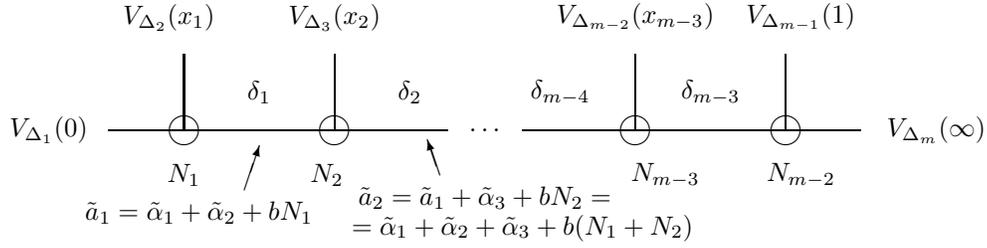

A natural generalization of relation (\ref{MainEquality}) is to
multipoint correlators
\begin{align*}
\Big< V_{\Delta_1}(z_1) \ \ldots \ V_{\Delta_m}(z_m) \Big>
\end{align*}
where $\Delta_i$'s are the dimensions of $m$ primary fields on a sphere.
With the help of $SL(2)$ transformations, one can always put the coordinates
$z_1, \ldots, z_m$ to positions $0, x_{1}, x_{2}, \ldots, x_{m-3}, 1, \infty$.
It is often convenient (for the reasons clarified below) to choose another parametrization
of coordinates:
\be
x_1 = q_1 q_2 \ldots q_{m-3}, \ \ \ \ \ \ x_2 = q_2 \ldots q_{m-3}, \ \ \ \ \ \
x_i= \prod\limits_{j = i}^{m-3} q_j, \ \ \ \ \ \ \ldots,\ \ \ \ \ \ x_{m-3} = q_{m-3}
\label{ChangeOfVariables}
\ee
Accordingly, in this parametrization the $m$-point
correlator on a sphere depends on $m - 3$ variables $q_i$
and can be written as a sum over $m - 3$ intermediate dimensions $\delta_i$:
{\fontsize{8pt}{0pt}
\be
\Big< \prod\limits_{i = 1}^{m-3} V_{\Delta_i}(x_i)
\cdot V_{\Delta_{m-2}}(0) V_{\Delta_{m-1}}(1)  V_{\Delta_m}(\infty) \Big>
= \sum\limits_{\delta_1 \ldots \delta_{m-3}}
C \big( \Delta_1, \Delta_2, \delta_1 \big) \prod\limits_{j = 1}^{m-2}  C \big( \delta_j, \Delta_{j+2}, \delta_{j+1} \big)
C \big( \delta_{m-3}, \Delta_{m-1}, \Delta_{m} \big) \times {\cal F}
\label{Multipoint1}
\ee}
with a shorthand notation
\begin{align}
\nonumber {\cal F} \ = \ & \prod\limits_{i = 1}^{m-3} q_i^{\delta_i - \Delta_1 - \ldots - \Delta_{i+1}}
\ B\big( \Delta_1, \ldots, \Delta_m, \delta_1, \ldots, \delta_{m-3}, \ c \
\big| \ q_1, \ldots, q_{m-3} \big) \times \\
& \ \ \ \times
\prod\limits_{i = 1}^{m-3} {\bar q}_i^{\bar\delta_i - \bar\Delta_{1} - \ldots - \bar\Delta_{i+1}}
\ {\overline B}\big( \bar\Delta_1, \ldots, \bar\Delta_m, \bar\delta_1, \ldots, \bar\delta_{m-3}, \ c \
\big| \ {\bar q}_1, \ldots, {\bar q}_{m-3} \big)
\label{Multipoint2}
\end{align}
As before, $C$'s are the 3-point functions
(which do not depend on $q$ and play the role of normalization constants for
$B(0, \ldots, 0) = 1$),
$c$ is the central charge and $B(q)$ is the m-point conformal block on a sphere.
The order of contractions of the 3-point functions in eq. (\ref{Multipoint1})
is most conveniently represented by a comb-like diagram
Fig.\ref{blockCircles}.

In analogy with the 4-point conformal block,
the functions $B\big( \Delta_1, \ldots, \Delta_m, \delta_1,
\ldots, \delta_{m-3}, \ c \ \big| \ q_1, \ldots, q_{m-3} \big)$
possess explicit series representations in \emph{positive} powers of $q_1, \ldots, q_{m-3}$,
which are described in
\cite{AlAnd2}. This is actually the reason to use the parametrization (\ref{ChangeOfVariables}):
in terms of the variables $x_1, \ldots, x_{m-3}$, the series would contain negative powers of variables,
which is less convenient. In this paper we are interested more in integral
representations for these conformal blocks, provided by the
multi-point Dotsenko-Fateev partition functions
$$Z_{DF} = \left< \prod_{a=0}^{m-2} :e^{\tilde\alpha_{a+1} \phi(x_a)}:\
 \prod_{a=1}^{m-2}\left( \int_0^{x_a} :e^{b\phi(z)}:\,dz\right)^{N_a} \right> = $$
 \vspace{-0.3cm}
\begin{align}
= \prod_{a<b} (x_b-x_a)^{\dfrac{\alpha_a\alpha_b}{2 \beta}}
\prod\limits_{a = 1}^{m - 2} \prod\limits_{i = 1}^{N_a}
\int\limits_{0}^{x_a} dz_{N_1 + \ldots + N_{a-1} + i}
\prod_{i < j} (z_j-z_i)^{\beta} \prod\limits_{ i }
\prod\limits_{a = 0}^{m-2} (z_i - x_a)^{\alpha_{a+1}}
\label{FDmpt}
\end{align}
where $\beta = b^2$, $\alpha_a = 2 b \tilde\alpha_a$ and $q_{0} = x_{0} = 0, q_{m-2}=x_{m-2}=1$.
It is natural to propose the following integral representation of spherical conformal blocks,
a multipoint counterpart of the relation (\ref{MainEquality}):
{\fontsize{8pt}{0pt}
\be
\boxed{ Z_{DF}\Big( \alpha_1, \ldots, \alpha_{m-1}, N_1, \ldots, N_{m-2}, \beta \ \Big|
\ q_1, \ldots, q_{m-3} \ \Big) = C_{DF} \cdot \prod\limits_{i = 1}^{m-3} q_i^{ {\rm deg}_i} \cdot
 B\big( \Delta_1, \ldots, \Delta_m, \delta_1, \ldots, \delta_{m-3},
\ c \ \big| \ q_1, \ldots, q_{m-3} \big)}
\label{MainEqualityFull}
\ee}
where ${\rm deg}_i = \delta_i - \Delta_1 - \ldots - \Delta_{i+1}$.
Again, $C_{DF}$ is the Dotsenko-Fateev normalization constant,
which does not depend on $q$ (but depends on all the other parameters)
and is a product of $m-2$ factors
\be
C_{DF} = C_{N_1}(\alpha_1, \alpha_2) C_{N_2}(\delta_1, \alpha_3)
\ldots C_{N_{m-2}}(\delta_{m-3}, \alpha_{m-1})
\ee
which are associated with the $m-2$ vertices of the diagram and depend on
respective screening multiplicities ($N_1, \ldots, N_{m-2}$) and incoming dimensions of
these vertices.
The relation between the $2m - 2$ Dotsenko-Fateev parameters $\alpha_1, \ldots, \alpha_{m-1}$,
$N_1, \ldots, N_{m-2}$, $\beta$ and the $2m - 2$ parameters of the conformal block
$\Delta_1, \ldots, \Delta_m$, $\delta_1, \ldots, \delta_{m-3}$, $c$ is
\begin{align}
\left\{
\begin{array}{lllll}
\Delta_i = \dfrac{ \alpha_i ( \alpha_i + 2 - 2 \beta )}{4 \beta}, \ \ i = 1, \ldots, {m-1}
\\
\\
\Delta_{m} = \dfrac{\big( \alpha_1 + \ldots + \alpha_{m-1} + 2 \beta ( N_1 + \ldots + N_{m-2} ) \big)
\big( \alpha_1 + \ldots + \alpha_{m-1} + 2 \beta ( N_1 + \ldots + N_{m-2} ) + 2 - 2 \beta \big) }{4\beta} \\
\\
\delta_i = \dfrac{\big( \alpha_1 + \ldots + \alpha_{i+1} + 2 \beta ( N_1 + \ldots + N_{i} ) \big)
\big( \alpha_1 + \ldots + \alpha_{i+1} + 2 \beta ( N_1 + \ldots + N_{i} ) + 2 - 2 \beta \big) }{4\beta}
\\
\\
c = 1 - 6 \left( \sqrt{\beta} - \dfrac{1}{\sqrt{\beta}} \right)^2 \\
\end{array}
\right.
\label{Relations1Multi}
\end{align}
or, in terms of free field variables,
\begin{align}
\left\{
\begin{array}{lllll}
\Delta_i = \tilde\alpha_i \left( \tilde\alpha_i + \dfrac{1}{b} - b \right), \ \ i = 1, \ldots, m-1
\\
\\
\Delta_{m} = \Big( \tilde\alpha_1 + \ldots + \tilde\alpha_{m-1} + b (N_1 + \ldots + N_{m-2}) \Big)
\left( \tilde\alpha_1 + \ldots + \tilde\alpha_{m-1} + b (N_1 + \ldots + N_{m-2}) + b - \dfrac{1}{b} \right) \\
\\
\delta_i = \Big( \tilde\alpha_1 + \ldots + \tilde\alpha_{i+1} + b (N_1 + \ldots + N_{i}) \Big)
 \left( \tilde\alpha_1 + \ldots + \tilde\alpha_{i+1} + b (N_1 + \ldots + N_{i}) + b - \dfrac{1}{b} \right)
\\
\\
c = 1 - 6 \left( b - \dfrac{1}{b} \right)^2 \ \\
\end{array}
\right.
\label{Relations2Multi}
\end{align}
The rules, which clearly stand behind these relations, are just the same as in the 4-point case,
see Fig.\ref{blockCircles}:

{\bf 1) Insertion of the first $m-3$ screening integrals with multiplicities $N_1, \ldots, N_{m-3}$
is needed to satisfy the free field
conservation law for the vertices,
$\tilde a_i = \tilde a_1+\ldots+\tilde a_{i+1} + bN_1 + \ldots + bN_i$ so that
the internal dimensions
$\delta_i = \tilde\gamma_i\left(\tilde\gamma_i + \frac{1}{b}-b\right)$
becomes unrelated to the free field values
$\Delta_{free} = (\tilde\alpha_1+\ldots+\tilde\alpha_{i+1})
\left(\tilde\alpha_1+\ldots+\tilde\alpha_{i+1}+\frac{1}{b}-b\right)$.
The $N_i$ integrals in these screenings are around positions of
$V_{\Delta_1}(0)$ and $V_{\Delta_{i+1}}(x_i)$.}

{\bf 2) Insertion of the last $N_{m-2}$ screening integrals is needed to satisfy
the free field conservation law for the most right vertex
$\tilde\alpha_m\cong \tilde a_{m-3}+\tilde\alpha_{m-1} + bN_{m-2}$.
}

{\bf 3) Alternatively, these additional $N_{m-2}$ screening integrals are needed to satisfy
the free field conservation law
\be
\sum_{i=1}^m \tilde\alpha_i + b \sum_{i=1}^{m-2} N_i = (g-1)\left(\frac{1}{b}-b\right)
\ee
($g$ is the genus, $g=0$ for the sphere),
by putting
$\tilde\alpha_{m} = - \tilde \alpha_1 - \ldots - \tilde\alpha_{m-1} - b(N_1+ \ldots + N_{m-3})
- \frac{1}{b} + b$.}

\bigskip

These rules naturally follow from the general logic of conformal field theory,
supplied by convincing evidence from direct calculations in the 4-point sector.
Of course, for $m > 4$ they still need to be carefully checked and/or rigorously proved.
Here we consider just the simplest check of (\ref{MainEqualityFull}) for the {\bf 5-point}
Dotsenko-Fateev integral:
\begin{align*}
Z_{DF} = q_1^{\dfrac{\alpha_1 (\alpha_2 + \alpha_3)}{2\beta}}
q_2^{\dfrac{\alpha_1 \alpha_2}{2\beta}}
(1 - q_1)^{\dfrac{\alpha_3 \alpha_4}{2\beta}}
(1 - q_1 q_2)^{\dfrac{\alpha_2 \alpha_4}{2\beta}}
(1 - q_2)^{\dfrac{\alpha_2 \alpha_3}{2\beta}} \times
\end{align*} \vspace{-0.3cm}
\begin{align}
\times \prod\limits_{i = 1}^{N_1} \int\limits_{0}^{q_1 q_2} d z_i \
\prod\limits_{i = 1}^{N_2} \int\limits_{0}^{q_2} d z_{N_1+i} \
\prod\limits_{i = 1}^{N_3} \int\limits_{0}^{1} d z_{N_1+N_2+i} \
\prod\limits_{i < j} (z_j - z_i)^{2 \beta}
\prod\limits_{i} z_i^{\alpha_1} (z_i - q_1 q_2)^{\alpha_2} (z_i - q_1)^{\alpha_3} (z_i - 1)^{\alpha_4}
\end{align}
The check of relation (\ref{MainEqualityFull}) for arbitrary values of
$\alpha_1, \ldots, \alpha_4$ is straightforward, but tedious.
To provide simple and clear evidence, we perform such a check in
the particular case of vanishing $\alpha_i$ and for $\beta = 1$.
In this case, the power series expansion of the Dotsenko-Fateev partition function
has the form
\begin{align}
Z_{DF} = q_1^{(N_1+N_2)^2} q_2^{N_2^2} \left[ 1 + \sum\limits_{k_1 + k_2 > 0}
J_{k_1,k_2} \Big( N_1, N_2, N_3 \Big) q_1^{k_1} q_2^{k_2} \right]
\end{align}
We find at level 1
\be
J_{1,0} \Big( N_1, N_2, N_3 \Big) =
-\dfrac{N_1 N_3 (N_1 + 2 N_2) (2 N_1+2 N_2+N_3)}{2(N_1+N_2)^2},
\ \ \ \ J_{0,1} \Big( N_1, N_2, N_3 \Big) = - \dfrac{N_1(N_1+2N_2)}{2}
\ee
and at level 2
$$
J_{2,0} \Big( N_1, N_2, N_3 \Big) =
\dfrac{N_1 N_3 (2 N_1 + 2 N_2 + N_3)(N_1 + 2 N_2)}
{(2 N_1 + 2 N_2 - 1)^2 (2 N_1 + 2 N_2)^2 (2 N_1 + 2 N_2 + 1)^2} \times$$
\begin{center}
$\times (-1 + 16 N_3 N_1^5 + 2 N_3 N_1 - 3 N_3^2 N_1^2-
6 N_3 N_1^3+8 N_3^2 N_1^4-8 N_2^3 N_3-4 N_2^2 N_3^2+
2 N_2 N_3-56 N_1 N_2^3-48 N_1^3 N_2-76 N_1^2 N_2^2+14 N_1 N_2+
32 N_2^4 N_3 N_1+16 N_2^3 N_3^2 N_1+112 N_2^3 N_3 N_1^2+
40 N_2^2 N_3^2 N_1^2+144 N_2^2 N_3 N_1^3-20 N_2^2 N_3 N_1-
18 N_2 N_3 N_1^2-6 N_2 N_3^2 N_1+32 N_2 N_3^2 N_1^3+
80 N_2 N_3 N_1^4+8 N_2^2+7 N_1^2+N_3^2-12 N_1^4-16 N_2^4)$
\end{center}
\be
J_{1,1} \Big( N_1, N_2, N_3 \Big) = \dfrac{N_3 (2 N_1 + 2 N_2 + N_3)
(-N_1^2-2 N_2^2-2 N_1 N_2+N_1^4+4 N_1^3 N_2+4 N_1^2 N_2^2)}
{(2 N_1 + 2 N_2)^2}
\ee
{\fontsize{9pt}{0pt}
\be
J_{0,2} \Big( N_1, N_2, N_3 \Big) =
\dfrac{N_1(N_1+2 N_2)(N_1^2-12 N_2^4+7 N_2^2-1+8 N_1^2 N_2^4
+16 N_1 N_2^5-3 N_1^2 N_2^2-6 N_1 N_2^3+2 N_1 N_2)}
{4(2 N_2-1)^2 (2 N_2+1)^2}
\ee}
These expressions need to be compared with the 5-point conformal block
\be
B\Big(\Delta_1, \Delta_2, \Delta_3, \Delta_4, \Delta_5, \delta_1, \delta_2, c \
| \ q_1, q_2 \ \Big) = 1 + \sum\limits_{k_1 + k_2 > 0}
B_{k_1,k_2} \Big( \Delta_1, \Delta_2, \Delta_3, \Delta_4, \Delta_5, \delta_1,
\delta_2, c\Big) q_1^{k_1} q_2^{k_2}
\ee
which was calculated explicitly in \cite{AlAnd2} at level 1:
\be
B_{10}\Big( \Delta_1, \Delta_2, \Delta_3, \Delta_4, \Delta_5, \delta_1, \delta_2, c\Big)
= \dfrac{(\delta_{1}+\Delta_{5}-\delta_{2}) (\delta_{1}+\Delta_{4}-\Delta_{1})}{2 \delta_{1}}
\ee
\be
B_{01}\Big( \Delta_1, \Delta_2, \Delta_3, \Delta_4, \Delta_5, \delta_1, \delta_2, c\Big)
= \dfrac{(\delta_{2}+\Delta_{5}-\delta_{1}) (\delta_{2}+\Delta_{2}-\Delta_{3})}{2 \delta_{2}}
\ee
and at level 2:
\begin{align}
& \nonumber B_{20}\Big( \Delta_1, \Delta_2, \Delta_3, \Delta_4, \Delta_5, \delta_1, \delta_2, c\Big)
= \dfrac{2 (\delta_{1}+2 \Delta_{5}-\delta_{2}) (\delta_{1}+2 \Delta_{4}-\Delta_{1})
(2 \delta_{1}+1)}{16 \delta_{1}^2-10 \delta_{1}+2 c \delta_{1}+c} - \\ \nonumber & \\
\nonumber &
- \dfrac{3(\delta_{1}+2 \Delta_{5}-\delta_{2}) (\delta_{1}+\Delta_{4}-\Delta_{1})
(\delta_{1}+\Delta_{4}-\Delta_{1}+1)}{16 \delta_{1}^2-10 \delta_{1}+2 c \delta_{1}+c} - \\
\nonumber & \\ \nonumber &
- \dfrac{3 (\delta_{2}-\Delta_{5}-\delta_{1}) (\delta_{2}-\Delta_{5}-\delta_{1}-1)
(\delta_{1}+2 \Delta_{4}-\Delta_{1})}{16 \delta_{1}^2-10 \delta_{1}+2 c \delta_{1}+c} + \\ \nonumber &
\\ &
+ \dfrac{(\delta_{2}-\Delta_{5}-\delta_{1}) (\delta_{2}-\Delta_{5}-\delta_{1}-1)
(\delta_{1}+\Delta_{4}-\Delta_{1}) (\delta_{1}+\Delta_{4}-\Delta_{1}+1) (8 \delta_{1}+c)}
{4\delta_{1}(16 \delta_{1}^2-10 \delta_{1}+2 c \delta_{1}+c)}
\end{align}
{\fontsize{9pt}{0pt}
\be
B_{11}\Big( \Delta_1, \Delta_2, \Delta_3, \Delta_4, \Delta_5, \delta_1, \delta_2, c\Big)
= \dfrac{((\delta_{1}+\Delta_{5}-\delta_{2}) (\delta_{2}+\Delta_{5}-\delta_{1}-1)+2 \delta_{1})
(\delta_{1}+\Delta_{4}-\Delta_{1}) (\delta_{2}+\Delta_{2}-\Delta_{3})}{4 \delta_{1} \delta_{2}}
\ee}
\begin{align}
& \nonumber B_{02}\Big( \Delta_1, \Delta_2, \Delta_3, \Delta_4, \Delta_5, \delta_1, \delta_2, c\Big)
= \dfrac{2 (\delta_{2}+2 \Delta_{5}-\delta_{1}) (2 \delta_{2}+1)(\delta_{2}+2 \Delta_{2}-\Delta_{3})}
{16 \delta_{2}^2-10 \delta_{2}+2 c \delta_{2}+c}
- \\ \nonumber & \\ \nonumber & - \dfrac{3 (\delta_{2}+2 \Delta_{5}-\delta_{1})
(\delta_{2}+\Delta_{2}-\Delta_{3}) (\delta_{2}+\Delta_{2}-\Delta_{3}+1)}
{16 \delta_{2}^2-10 \delta_{2}+2 c \delta_{2}+c}
- \\ \nonumber & \\ \nonumber & - \dfrac{3 (\delta_{2}+\Delta_{5}-\delta_{1})
(\delta_{2}+\Delta_{5}-\delta_{1}+1) (\delta_{2}+2 \Delta_{2}-\Delta_{3})}
{16 \delta_{2}^2-10 \delta_{2}+2 c \delta_{2}+c}
+ \\ \nonumber & \\ & +
\dfrac{(\delta_{2}+\Delta_{5}-\delta_{1}) (\delta_{2}+\Delta_{5}-\delta_{1}+1)
(8 \delta_{2}+c) (\delta_{2}+\Delta_{2}-\Delta_{3}) (\delta_{2}+\Delta_{2}-\Delta_{3}+1)}
{4 \delta_{2} (16 \delta_{2}^2-10 \delta_{2}+2 c \delta_{2}+c)}
\end{align}
Comparing these expressions, one finds
\be
J_{k_1,k_2} \Big( N_1, N_2, N_3 \Big) = B_{k_1,k_2}\Big( 0,0,0,0, (N_1+N_2+N_3)^2, (N_1+N_2)^2, (N_2)^2, c = 1\Big)
\ee
at least, for $(k_1,k_2) = (1,0),(0,1),(2,0),(1,1),(0,2)$. This relation can be considered
as an evidence
that Dotsenko-Fateev integrals continue to provide a relevant description
of spherical (genus zero) conformal blocks beyond four primaries.

\section{Dotsenko-Fateev integral on a torus
\label{tor}}

Since the Dotsenko-Fateev integrals (\ref{FD4pt}) and (\ref{FDmpt})
are made from the correlators of free
fields, one can easily consider their generalizations on higher
genus Riemann surfaces.
There is no an evident way to associate these integrals with the corresponding
conformal blocks, and we shall now see that, indeed, the most naive attempt
leads to an expression which is surprisingly similar, but still different.
Thus generalization of the AGT conjecture in this direction still remains to be found.
This short subsection only describes the setting.

If a Riemann sphere is substituted by a torus, then at the r.h.s. of (\ref{FD4pt})
and (\ref{FDmpt}) the role of the holomorphic Green functions is played
by the odd theta-functions
\begin{align}
\theta_*(x) \sim \sin\frac{x}{2} - q \sin\frac{3x}{2}
+ q^3 \sin\frac{5x}{2} - q^6 \sin\frac{7x}{2} + \ldots =
\sum\limits_{n = 0}^{\infty} (-1)^n q^{n(n+1)/2} \sin\frac{(2n+1)x}{2}
\end{align}
and the typical Dotsenko-Fateev integral looks like
\begin{align}
Z_{DF}(\alpha,\beta,N,q) = \int\limits_{0}^{2\pi} dz_1 \ldots
\int\limits_{0}^{2\pi} dz_N \prod\limits_{i < j} \theta_*(z_i - z_j)^{2\beta}
\prod\limits_{i} \theta_*(z_i)^{\alpha} = {\rm const} \cdot
( 1 + J_1 q + J_2 q^2 + \ldots )
\label{Ztor}
\end{align}
There are many questions to ask about the r.h.s. of this formula, to mention
just a few: the integrand is not double periodic;
the conservation law $\tilde\alpha + Nb=0$ is not satisfied
(and there is no infinitely remote point to hide the compensating insertion at);
the integral is taken only along the $A$-cycle;
it is unclear if the argument of \cite{KMMMP}, allowing to neglect the
second set of screening charges, is applicable and so on.
Not to repeat that there are no basic reasons to identify this integral
or any of its modifications with the toric conformal block,
except for a certain similarity to the AGT conjecture in the form of
(\ref{MainEquality}) and (\ref{MainEqualityFull}).

For good or for bad, the coefficients at the r.h.s. of (\ref{Ztor})
can be straightforwardly calculated. To do this, notice that the Green function can be
represented as a product
\be
\theta_*(x) \sim \sin\frac{x}{2}\
\left\{ 1 + \left(1 - 4 \cos^2 \frac{x}{2}\right)q + \left( 1 - 12 \cos^2 \frac{x}{2}
+ 16 \cos^4 \frac{x}{2} \right) q^2 + \ldots \right\}
\ee
Using this formula, it is easy to find
\be
J_1 = \alpha \sum\limits_{i = 1}^{N} \Big< 1 - 4 \cos^2 \dfrac{z_i}{2} \Big> + 2 \beta \sum\limits_{i < j} \Big< 1 - 4 \cos^2 \dfrac{z_i - z_j}{2} \Big>
\ee
where the Dotsenko-Fateev correlators are defined as
\be
\Big< f(z_1, \ldots, z_N) \Big> = \dfrac{ \int\limits_{0}^{2\pi} dz_1 \ldots
\int\limits_{0}^{2\pi} dz_N \ f(z_1, \ldots, z_N) \ \prod\limits_{i < j} \left( \sin \dfrac{z_i - z_j}{2} \right)^{2\beta}
\prod\limits_{i} \left( \sin \dfrac{z_i}{2} \right)^{\alpha} }{ \int\limits_{0}^{2\pi} dz_1 \ldots
\int\limits_{0}^{2\pi} dz_N \prod\limits_{i < j} \left( \sin \dfrac{z_i - z_j}{2} \right)^{2\beta}
\prod\limits_{i} \left( \sin \dfrac{z_i}{2} \right)^{\alpha} }
\ee
Direct calculations give
\be
\Big< \cos^2 \dfrac{z_i}{2} \Big> = \dfrac{1 + (N-1) \beta}
{2 ((N-1) \beta + \alpha/2 + 1)}
\ee
and
\be
\Big< \cos^2 \dfrac{z_i - z_j}{2} \Big> =
\dfrac{(N-1)(N-2) \beta^2 + \beta \Big((N-2) \alpha + (2 N-3)\Big) +
(1 +  \alpha + \alpha^2/2)}{2 \Big((N-1) \beta + \alpha/2 + 1\Big)^2}
\ee
Note that the r.h.s. does not depend on $i, j$
because of the permutation symmetry. Consequently,
\be
J_1 = \alpha N \Big< 1 - 4 \cos(z_1)^2 \Big> + \beta N(N-1)
\Big< 1 - 4 \cos(z_1 - z_2)^2 \Big> =
\label{J1tor}
\ee
\vspace{-0.5cm}
{\fontsize{9pt}{0pt}
\begin{align}
\emph{}= - \dfrac{N \beta (N-1) (\beta N-\beta+1) (\beta N-3 \beta+1)
+N (3 \beta N-3 \beta+2 \beta^2 N^2+1+4 \beta^2-6 \beta^2 N) \alpha
+\frac{3}{4} N \beta (N-1) \alpha^2-\frac{1}{4} N \alpha^3}
{\left(1 - \beta + \frac{\alpha}{2} + \beta N\right)^2}
\nn
\end{align}}
The integral $Z_{DF}$ needs to be compared with the toric 1-point function
\be
B = 1 + B_1 q + B_2 q^2 + \ldots
\label{Btor}
\ee
where (see, e.g., \cite{Pog,AlAnd1}),
\be
B_1 = \frac{\Delta^2_{ext}-\Delta_{ext} + 2\Delta}{\Delta}
\ee
Since $\Delta$ and $\Delta_{ext}$ are given by formulas like (\ref{Relations2})
\be
\Delta = \dfrac{a( a + 2 - 2 \beta)}{4 \beta}, \ \ \ \ \Delta_{ext}
= \dfrac{\alpha( \alpha + 2 - 2 \beta)}{4 \beta}
\ee
which are not full squares, it is clear already from a look at the
denominators that (\ref{J1tor}) does not match (\ref{Btor}).
At the same time the difference is not quite as drastic as it could be.
It looks like the most severe discrepancy between $Z_{DF}$ and the toric
conformal block is just in the number of free parameters. In the Dotsenko-Fateev integral,
$\alpha$ is actually fixed by the free field conservation law
\be
\alpha + 2 \beta N = 0
\ee
The most straightforward way to relax this restriction is to introduce $N_2$
additional integrals over the $B$-cycle.
In terms of variables $z_i$, this would result
in the following modification of the Dotsenko-Fateev integral:
\be
Z_{DF}(\alpha,\beta,N_1,N_2,q) = \int\limits_{0}^{2\pi} dz_1 \ldots
\int\limits_{0}^{2\pi} dz_{N_1} \int_B dz_{N_1+1} \ldots
\int_B dz_{N_1+N_2} \prod\limits_{i < j} \theta_*(z_i - z_j)^{2\beta}
\prod\limits_{i} \theta_*(z_i)^{\alpha}
\ee
and the new conservation law would take form
\be
\alpha + 2 \beta (N_1 + N_2) = 0
\label{cola}
\ee
There is also a selection rule
which states
\be
a + \alpha + 2 \beta N_1 = 2 \beta - 2 - a
\ee
in accordance with
\be
\Delta[a] = \dfrac{a( a + 2 - 2 \beta)}{4 \beta} = \Delta[2\beta - 2 -\alpha]
\ee
Thus
\be
a = (N_2+1)\beta-1
\label{abe}
\ee

The exact formula (\ref{J1tor}) for $J_1$ is sufficient
to describe only the case when $N_1 = N$ and  $N_2 = 0$.
In this situation, one gets from (\ref{abe}) and (\ref{cola})
\be
a = \beta-1, \ \ \ \ \ \alpha=-2\beta N
\ee
so that $\Delta$ and $\Delta_{ext}$ are equal to
\be
\Delta = -\dfrac{(\beta - 1)^2}{4\beta}, \ \ \ \ \Delta_{ext} = N(\beta N + \beta - 1)
\ee
Substituting these expressions into $B_1$ and $J_1$, one obtains
\be
B_1 = \dfrac{2 \beta^3 N^4-4 \beta^2 N^3+4 \beta^3 N^3-6 \beta^2 N^2+2 \beta^3 N^2
-2 \beta^2 N+2 \beta N^2+2 \beta N-2 \beta+\beta^2+1}{(\beta-1)^2}
\ee
and
\be
J_1 = - \dfrac{N \beta (N+1) (2 \beta^2 N^2-4 \beta N+2 \beta^2 N-1+4 \beta-3 \beta^2)}{(\beta-1)^2}
= B_1 + 3 \Delta_{ext} - 1 + 3N
\ee
As one can see, the discrepancy between these two quantities is quite moderate.
In principle, the extra terms $3 \Delta_{ext} - 1 + 3N$ can be easily absorbed into
$U(1)$ factors,
but it does not seem natural to introduce $U(1)$ factors which depend explicitly not only on
dimensions, but also on $N$. A straightforward solution may be to divide each screening
integral by $(1-q)^3$ or by $q^{-1/8}\eta(q)^3$, where
\be
\eta(q) = q^{1/24}\prod\limits_{n = 1}^{\infty} (1 - q^n)
\ee
is the Dedekind eta-function.
However, it remains to be checked if such a prescription can correctly
reproduce the toric conformal block at level 2 and at higher levels.
The problem clearly deserves further investigation.
As already explained in s.\ref{prob}, an even bigger problem is to
adequately switch on $N_2\neq 0$.

\section{Conclusion}

In this paper, we provide some further evidence in support of the modified
AGT conjecture \cite{DVagt,MMSh1},
identifying conformal blocks with matrix integrals in the DV phase,
and making no any direct reference to the Nekrasov functions.
This evidence, together with the earlier results in \cite{Ito,Egu,Wilma,FHKT},
seems to be sufficient to establish the relation at the spherical level.
As a byproduct, we found explicit formulas for the analytical continuation
of the Dotsenko-Fateev integrals, which appear parallel to the formulas
\cite{DVph} for the CIV superpotential.
However, nothing like a straightforward proof of the modified AGT conjecture
is yet available, and it can not yet be used to prove neither the original
AGT hypothesis \cite{AGT1} along the lines of \cite{DVagt,MMSh1},
nor its BS/SW version \cite{GKMMM,NSh,MMBS}.
Moreover, generalization to higher genera also runs into certain problems,
briefly described in the.s.\ref{prob} above.
The future proof should clarify these subtle points
and establish a direct relation between the decomposition
formulas \cite{deco} for matrix model partition functions
and the conformal block expansions \cite{CFT,AGT1,MMMagt}
into the triple vertices and the inverse Shapovalov matrices.

\newpage

\section*{Appendix. Explicit expression for
$J_2(\alpha_1,\alpha_2,\alpha_3,N_1,N_2,\beta = 1)$}

$$ J_2(\alpha_1,\alpha_2,\alpha_3,N_1,N_2,\beta = 1) \ = \
\dfrac{1}{(\alpha_{1}+\alpha_{2}+2 N_{1}+1)^2
(2 N_{1}+\alpha_{1}+\alpha_{2})^2(\alpha_{1}+\alpha_{2}+2 N_{1}-1)^2} \
\times \emph{} $$
\begin{center}
$ \Big( -2 \alpha_{2}^3 \alpha_{3}-4 \alpha_{1} \alpha_{2}^2
\alpha_{3}-2 \alpha_{1}^2 \alpha_{2} \alpha_{3}-4 N_{2} \alpha_{2}^2
\alpha_{3}-4 N_{2} \alpha_{2}^3-4 N_{2} \alpha_{1} \alpha_{2}
\alpha_{3}-8 N_{2} \alpha_{1} \alpha_{2}^2-4 N_{2} \alpha_{1}^2
\alpha_{2}-4 N_{2}^2 \alpha_{2}^2-4 N_{2}^2 \alpha_{1} \alpha_{2}-8
N_{1} \alpha_{2}^2 \alpha_{3}-12 N_{1} \alpha_{1} \alpha_{2}
\alpha_{3}-4 N_{1} \alpha_{1}^2 \alpha_{3}-8 N_{1} N_{2} \alpha_{2}
\alpha_{3}-16 N_{1} N_{2} \alpha_{2}^2-8 N_{1} N_{2} \alpha_{1}
\alpha_{3}-24 N_{1} N_{2} \alpha_{1} \alpha_{2}-8 N_{1} N_{2}
\alpha_{1}^2-8 N_{1} N_{2}^2 \alpha_{2}-8 N_{1} N_{2}^2
\alpha_{1}-12 N_{1}^2 \alpha_{2} \alpha_{3}-12 N_{1}^2 \alpha_{1}
\alpha_{3}-8 N_{1}^2 N_{2} \alpha_{3}-24 N_{1}^2 N_{2} \alpha_{2}-24
N_{1}^2 N_{2} \alpha_{1}-8 N_{1}^2 N_{2}^2-8 N_{1}^3 \alpha_{3}-16
N_{1}^3 N_{2}+\alpha_{2}^4 \alpha_{3}^2+4 \alpha_{2}^5 \alpha_{3}+2
\alpha_{1} \alpha_{2}^3 \alpha_{3}^2+16 \alpha_{1} \alpha_{2}^4
\alpha_{3}+\alpha_{1}^2 \alpha_{2}^2 \alpha_{3}^2+24 \alpha_{1}^2
\alpha_{2}^3 \alpha_{3}+16 \alpha_{1}^3 \alpha_{2}^2 \alpha_{3}+4
\alpha_{1}^4 \alpha_{2} \alpha_{3}+4 N_{2} \alpha_{2}^3
\alpha_{3}^2+12 N_{2} \alpha_{2}^4 \alpha_{3}+8 N_{2} \alpha_{2}^5+8
N_{2} \alpha_{1} \alpha_{2}^2 \alpha_{3}^2+36 N_{2} \alpha_{1}
\alpha_{2}^3 \alpha_{3}+32 N_{2} \alpha_{1} \alpha_{2}^4+4 N_{2}
\alpha_{1}^2 \alpha_{2} \alpha_{3}^2+36 N_{2} \alpha_{1}^2
\alpha_{2}^2 \alpha_{3}+48 N_{2} \alpha_{1}^2 \alpha_{2}^3+12 N_{2}
\alpha_{1}^3 \alpha_{2} \alpha_{3}+32 N_{2} \alpha_{1}^3
\alpha_{2}^2+8 N_{2} \alpha_{1}^4 \alpha_{2}+4 N_{2}^2 \alpha_{2}^2
\alpha_{3}^2+12 N_{2}^2 \alpha_{2}^3 \alpha_{3}+12 N_{2}^2
\alpha_{2}^4+4 N_{2}^2 \alpha_{1} \alpha_{2} \alpha_{3}^2+24 N_{2}^2
\alpha_{1} \alpha_{2}^2 \alpha_{3}+36 N_{2}^2 \alpha_{1}
\alpha_{2}^3+12 N_{2}^2 \alpha_{1}^2 \alpha_{2} \alpha_{3}+36
N_{2}^2 \alpha_{1}^2 \alpha_{2}^2+12 N_{2}^2 \alpha_{1}^3
\alpha_{2}+8 N_{2}^3 \alpha_{2}^2 \alpha_{3}+8 N_{2}^3
\alpha_{2}^3+8 N_{2}^3 \alpha_{1} \alpha_{2} \alpha_{3}+16 N_{2}^3
\alpha_{1} \alpha_{2}^2+8 N_{2}^3 \alpha_{1}^2 \alpha_{2}+4 N_{2}^4
\alpha_{2}^2+4 N_{2}^4 \alpha_{1} \alpha_{2}+4 N_{1} \alpha_{2}^3
\alpha_{3}^2+32 N_{1} \alpha_{2}^4 \alpha_{3}+4 N_{1} \alpha_{1}
\alpha_{2}^2 \alpha_{3}^2+100 N_{1} \alpha_{1} \alpha_{2}^3
\alpha_{3}+112 N_{1} \alpha_{1}^2 \alpha_{2}^2 \alpha_{3}+52 N_{1}
\alpha_{1}^3 \alpha_{2} \alpha_{3}+8 N_{1} \alpha_{1}^4
\alpha_{3}+16 N_{1} N_{2} \alpha_{2}^2 \alpha_{3}^2+72 N_{1} N_{2}
\alpha_{2}^3 \alpha_{3}+64 N_{1} N_{2} \alpha_{2}^4+24 N_{1} N_{2}
\alpha_{1} \alpha_{2} \alpha_{3}^2+160 N_{1} N_{2} \alpha_{1}
\alpha_{2}^2 \alpha_{3}+200 N_{1} N_{2} \alpha_{1} \alpha_{2}^3+8
N_{1} N_{2} \alpha_{1}^2 \alpha_{3}^2+112 N_{1} N_{2} \alpha_{1}^2
\alpha_{2} \alpha_{3}+224 N_{1} N_{2} \alpha_{1}^2 \alpha_{2}^2+24
N_{1} N_{2} \alpha_{1}^3 \alpha_{3}+104 N_{1} N_{2} \alpha_{1}^3
\alpha_{2}+16 N_{1} N_{2} \alpha_{1}^4+8 N_{1} N_{2}^2 \alpha_{2}
\alpha_{3}^2+48 N_{1} N_{2}^2 \alpha_{2}^2 \alpha_{3}+72 N_{1}
N_{2}^2 \alpha_{2}^3+8 N_{1} N_{2}^2 \alpha_{1} \alpha_{3}^2+72
N_{1} N_{2}^2 \alpha_{1} \alpha_{2} \alpha_{3}+160 N_{1} N_{2}^2
\alpha_{1} \alpha_{2}^2+24 N_{1} N_{2}^2 \alpha_{1}^2 \alpha_{3}+112
N_{1} N_{2}^2 \alpha_{1}^2 \alpha_{2}+24 N_{1} N_{2}^2
\alpha_{1}^3+16 N_{1} N_{2}^3 \alpha_{2} \alpha_{3}+32 N_{1} N_{2}^3
\alpha_{2}^2+16 N_{1} N_{2}^3 \alpha_{1} \alpha_{3}+48 N_{1} N_{2}^3
\alpha_{1} \alpha_{2}+16 N_{1} N_{2}^3 \alpha_{1}^2+8 N_{1} N_{2}^4
\alpha_{2}+8 N_{1} N_{2}^4 \alpha_{1}+4 N_{1}^2 \alpha_{2}^2
\alpha_{3}^2+100 N_{1}^2 \alpha_{2}^3 \alpha_{3}+240 N_{1}^2
\alpha_{1} \alpha_{2}^2 \alpha_{3}+192 N_{1}^2 \alpha_{1}^2
\alpha_{2} \alpha_{3}+52 N_{1}^2 \alpha_{1}^3 \alpha_{3}+24 N_{1}^2
N_{2} \alpha_{2} \alpha_{3}^2+160 N_{1}^2 N_{2} \alpha_{2}^2
\alpha_{3}+200 N_{1}^2 N_{2} \alpha_{2}^3+24 N_{1}^2 N_{2}
\alpha_{1} \alpha_{3}^2+264 N_{1}^2 N_{2} \alpha_{1} \alpha_{2}
\alpha_{3}+480 N_{1}^2 N_{2} \alpha_{1} \alpha_{2}^2+112 N_{1}^2
N_{2} \alpha_{1}^2 \alpha_{3}+384 N_{1}^2 N_{2} \alpha_{1}^2
\alpha_{2}+104 N_{1}^2 N_{2} \alpha_{1}^3+8 N_{1}^2 N_{2}^2
\alpha_{3}^2+72 N_{1}^2 N_{2}^2 \alpha_{2} \alpha_{3}+160 N_{1}^2
N_{2}^2 \alpha_{2}^2+72 N_{1}^2 N_{2}^2 \alpha_{1} \alpha_{3}+264
N_{1}^2 N_{2}^2 \alpha_{1} \alpha_{2}+112 N_{1}^2 N_{2}^2
\alpha_{1}^2+16 N_{1}^2 N_{2}^3 \alpha_{3}+48 N_{1}^2 N_{2}^3
\alpha_{2}+48 N_{1}^2 N_{2}^3 \alpha_{1}+8 N_{1}^2 N_{2}^4+160
N_{1}^3 \alpha_{2}^2 \alpha_{3}+280 N_{1}^3 \alpha_{1} \alpha_{2}
\alpha_{3}+128 N_{1}^3 \alpha_{1}^2 \alpha_{3}+16 N_{1}^3 N_{2}
\alpha_{3}^2+176 N_{1}^3 N_{2} \alpha_{2} \alpha_{3}+320 N_{1}^3
N_{2} \alpha_{2}^2+176 N_{1}^3 N_{2} \alpha_{1} \alpha_{3}+560
N_{1}^3 N_{2} \alpha_{1} \alpha_{2}+256 N_{1}^3 N_{2}
\alpha_{1}^2+48 N_{1}^3 N_{2}^2 \alpha_{3}+176 N_{1}^3 N_{2}^2
\alpha_{2}+176 N_{1}^3 N_{2}^2 \alpha_{1}+32 N_{1}^3 N_{2}^3+140
N_{1}^4 \alpha_{2} \alpha_{3}+140 N_{1}^4 \alpha_{1} \alpha_{3}+88
N_{1}^4 N_{2} \alpha_{3}+280 N_{1}^4 N_{2} \alpha_{2}+280 N_{1}^4
N_{2} \alpha_{1}+88 N_{1}^4 N_{2}^2+56 N_{1}^5 \alpha_{3}+112
N_{1}^5 N_{2}-2 \alpha_{2}^6 \alpha_{3}^2-2 \alpha_{2}^7
\alpha_{3}-8 \alpha_{1} \alpha_{2}^5 \alpha_{3}^2-12 \alpha_{1}
\alpha_{2}^6 \alpha_{3}-12 \alpha_{1}^2 \alpha_{2}^4 \alpha_{3}^2-30
\alpha_{1}^2 \alpha_{2}^5 \alpha_{3}-8 \alpha_{1}^3 \alpha_{2}^3
\alpha_{3}^2-40 \alpha_{1}^3 \alpha_{2}^4 \alpha_{3}-2 \alpha_{1}^4
\alpha_{2}^2 \alpha_{3}^2-30 \alpha_{1}^4 \alpha_{2}^3 \alpha_{3}-12
\alpha_{1}^5 \alpha_{2}^2 \alpha_{3}-2 \alpha_{1}^6 \alpha_{2}
\alpha_{3}-8 N_{2} \alpha_{2}^5 \alpha_{3}^2-12 N_{2} \alpha_{2}^6
\alpha_{3}-4 N_{2} \alpha_{2}^7-28 N_{2} \alpha_{1} \alpha_{2}^4
\alpha_{3}^2-56 N_{2} \alpha_{1} \alpha_{2}^5 \alpha_{3}-24 N_{2}
\alpha_{1} \alpha_{2}^6-36 N_{2} \alpha_{1}^2 \alpha_{2}^3
\alpha_{3}^2-104 N_{2} \alpha_{1}^2 \alpha_{2}^4 \alpha_{3}-60 N_{2}
\alpha_{1}^2 \alpha_{2}^5-20 N_{2} \alpha_{1}^3 \alpha_{2}^2
\alpha_{3}^2-96 N_{2} \alpha_{1}^3 \alpha_{2}^3 \alpha_{3}-80 N_{2}
\alpha_{1}^3 \alpha_{2}^4-4 N_{2} \alpha_{1}^4 \alpha_{2}
\alpha_{3}^2-44 N_{2} \alpha_{1}^4 \alpha_{2}^2 \alpha_{3}-60 N_{2}
\alpha_{1}^4 \alpha_{2}^3-8 N_{2} \alpha_{1}^5 \alpha_{2}
\alpha_{3}-24 N_{2} \alpha_{1}^5 \alpha_{2}^2-4 N_{2} \alpha_{1}^6
\alpha_{2}-8 N_{2}^2 \alpha_{2}^4 \alpha_{3}^2-24 N_{2}^2
\alpha_{2}^5 \alpha_{3}-12 N_{2}^2 \alpha_{2}^6-20 N_{2}^2
\alpha_{1} \alpha_{2}^3 \alpha_{3}^2-84 N_{2}^2 \alpha_{1}
\alpha_{2}^4 \alpha_{3}-56 N_{2}^2 \alpha_{1} \alpha_{2}^5-16
N_{2}^2 \alpha_{1}^2 \alpha_{2}^2 \alpha_{3}^2-108 N_{2}^2
\alpha_{1}^2 \alpha_{2}^3 \alpha_{3}-104 N_{2}^2 \alpha_{1}^2
\alpha_{2}^4-4 N_{2}^2 \alpha_{1}^3 \alpha_{2} \alpha_{3}^2-60
N_{2}^2 \alpha_{1}^3 \alpha_{2}^2 \alpha_{3}-96 N_{2}^2 \alpha_{1}^3
\alpha_{2}^3-12 N_{2}^2 \alpha_{1}^4 \alpha_{2} \alpha_{3}-44
N_{2}^2 \alpha_{1}^4 \alpha_{2}^2-8 N_{2}^2 \alpha_{1}^5
\alpha_{2}-16 N_{2}^3 \alpha_{2}^4 \alpha_{3}-16 N_{2}^3
\alpha_{2}^5-40 N_{2}^3 \alpha_{1} \alpha_{2}^3 \alpha_{3}-56
N_{2}^3 \alpha_{1} \alpha_{2}^4-32 N_{2}^3 \alpha_{1}^2 \alpha_{2}^2
\alpha_{3}-72 N_{2}^3 \alpha_{1}^2 \alpha_{2}^3-8 N_{2}^3
\alpha_{1}^3 \alpha_{2} \alpha_{3}-40 N_{2}^3 \alpha_{1}^3
\alpha_{2}^2-8 N_{2}^3 \alpha_{1}^4 \alpha_{2}-8 N_{2}^4
\alpha_{2}^4-20 N_{2}^4 \alpha_{1} \alpha_{2}^3-16 N_{2}^4
\alpha_{1}^2 \alpha_{2}^2-4 N_{2}^4 \alpha_{1}^3 \alpha_{2}-16 N_{1}
\alpha_{2}^5 \alpha_{3}^2-24 N_{1} \alpha_{2}^6 \alpha_{3}-52 N_{1}
\alpha_{1} \alpha_{2}^4 \alpha_{3}^2-120 N_{1} \alpha_{1}
\alpha_{2}^5 \alpha_{3}-60 N_{1} \alpha_{1}^2 \alpha_{2}^3
\alpha_{3}^2-244 N_{1} \alpha_{1}^2 \alpha_{2}^4 \alpha_{3}-28 N_{1}
\alpha_{1}^3 \alpha_{2}^2 \alpha_{3}^2-256 N_{1} \alpha_{1}^3
\alpha_{2}^3 \alpha_{3}-4 N_{1} \alpha_{1}^4 \alpha_{2}
\alpha_{3}^2-144 N_{1} \alpha_{1}^4 \alpha_{2}^2 \alpha_{3}-40 N_{1}
\alpha_{1}^5 \alpha_{2} \alpha_{3}-4 N_{1} \alpha_{1}^6
\alpha_{3}-56 N_{1} N_{2} \alpha_{2}^4 \alpha_{3}^2-112 N_{1} N_{2}
\alpha_{2}^5 \alpha_{3}-48 N_{1} N_{2} \alpha_{2}^6-144 N_{1} N_{2}
\alpha_{1} \alpha_{2}^3 \alpha_{3}^2-416 N_{1} N_{2} \alpha_{1}
\alpha_{2}^4 \alpha_{3}-240 N_{1} N_{2} \alpha_{1} \alpha_{2}^5-128
N_{1} N_{2} \alpha_{1}^2 \alpha_{2}^2 \alpha_{3}^2-592 N_{1} N_{2}
\alpha_{1}^2 \alpha_{2}^3 \alpha_{3}-488 N_{1} N_{2} \alpha_{1}^2
\alpha_{2}^4-48 N_{1} N_{2} \alpha_{1}^3 \alpha_{2} \alpha_{3}^2-400
N_{1} N_{2} \alpha_{1}^3 \alpha_{2}^2 \alpha_{3}-512 N_{1} N_{2}
\alpha_{1}^3 \alpha_{2}^3-8 N_{1} N_{2} \alpha_{1}^4
\alpha_{3}^2-128 N_{1} N_{2} \alpha_{1}^4 \alpha_{2} \alpha_{3}-288
N_{1} N_{2} \alpha_{1}^4 \alpha_{2}^2-16 N_{1} N_{2} \alpha_{1}^5
\alpha_{3}-80 N_{1} N_{2} \alpha_{1}^5 \alpha_{2}-8 N_{1} N_{2}
\alpha_{1}^6-40 N_{1} N_{2}^2 \alpha_{2}^3 \alpha_{3}^2-168 N_{1}
N_{2}^2 \alpha_{2}^4 \alpha_{3}-112 N_{1} N_{2}^2 \alpha_{2}^5-64
N_{1} N_{2}^2 \alpha_{1} \alpha_{2}^2 \alpha_{3}^2-432 N_{1} N_{2}^2
\alpha_{1} \alpha_{2}^3 \alpha_{3}-416 N_{1} N_{2}^2 \alpha_{1}
\alpha_{2}^4-32 N_{1} N_{2}^2 \alpha_{1}^2 \alpha_{2}
\alpha_{3}^2-384 N_{1} N_{2}^2 \alpha_{1}^2 \alpha_{2}^2
\alpha_{3}-592 N_{1} N_{2}^2 \alpha_{1}^2 \alpha_{2}^3-8 N_{1}
N_{2}^2 \alpha_{1}^3 \alpha_{3}^2-144 N_{1} N_{2}^2 \alpha_{1}^3
\alpha_{2} \alpha_{3}-400 N_{1} N_{2}^2 \alpha_{1}^3 \alpha_{2}^2-24
N_{1} N_{2}^2 \alpha_{1}^4 \alpha_{3}-128 N_{1} N_{2}^2 \alpha_{1}^4
\alpha_{2}-16 N_{1} N_{2}^2 \alpha_{1}^5-80 N_{1} N_{2}^3
\alpha_{2}^3 \alpha_{3}-112 N_{1} N_{2}^3 \alpha_{2}^4-128 N_{1}
N_{2}^3 \alpha_{1} \alpha_{2}^2 \alpha_{3}-288 N_{1} N_{2}^3
\alpha_{1} \alpha_{2}^3-64 N_{1} N_{2}^3 \alpha_{1}^2 \alpha_{2}
\alpha_{3}-256 N_{1} N_{2}^3 \alpha_{1}^2 \alpha_{2}^2-16 N_{1}
N_{2}^3 \alpha_{1}^3 \alpha_{3}-96 N_{1} N_{2}^3 \alpha_{1}^3
\alpha_{2}-16 N_{1} N_{2}^3 \alpha_{1}^4-40 N_{1} N_{2}^4
\alpha_{2}^3-64 N_{1} N_{2}^4 \alpha_{1} \alpha_{2}^2-32 N_{1}
N_{2}^4 \alpha_{1}^2 \alpha_{2}-8 N_{1} N_{2}^4 \alpha_{1}^3-52
N_{1}^2 \alpha_{2}^4 \alpha_{3}^2-120 N_{1}^2 \alpha_{2}^5
\alpha_{3}-132 N_{1}^2 \alpha_{1} \alpha_{2}^3 \alpha_{3}^2-492
N_{1}^2 \alpha_{1} \alpha_{2}^4 \alpha_{3}-112 N_{1}^2 \alpha_{1}^2
\alpha_{2}^2 \alpha_{3}^2-796 N_{1}^2 \alpha_{1}^2 \alpha_{2}^3
\alpha_{3}-36 N_{1}^2 \alpha_{1}^3 \alpha_{2} \alpha_{3}^2-636
N_{1}^2 \alpha_{1}^3 \alpha_{2}^2 \alpha_{3}-4 N_{1}^2 \alpha_{1}^4
\alpha_{3}^2-252 N_{1}^2 \alpha_{1}^4 \alpha_{2} \alpha_{3}-40
N_{1}^2 \alpha_{1}^5 \alpha_{3}-144 N_{1}^2 N_{2} \alpha_{2}^3
\alpha_{3}^2-416 N_{1}^2 N_{2} \alpha_{2}^4 \alpha_{3}-240 N_{1}^2
N_{2} \alpha_{2}^5-264 N_{1}^2 N_{2} \alpha_{1} \alpha_{2}^2
\alpha_{3}^2-1200 N_{1}^2 N_{2} \alpha_{1} \alpha_{2}^3
\alpha_{3}-984 N_{1}^2 N_{2} \alpha_{1} \alpha_{2}^4-168 N_{1}^2
N_{2} \alpha_{1}^2 \alpha_{2} \alpha_{3}^2-1280 N_{1}^2 N_{2}
\alpha_{1}^2 \alpha_{2}^2 \alpha_{3}-1592 N_{1}^2 N_{2} \alpha_{1}^2
\alpha_{2}^3-48 N_{1}^2 N_{2} \alpha_{1}^3 \alpha_{3}^2-624 N_{1}^2
N_{2} \alpha_{1}^3 \alpha_{2} \alpha_{3}-1272 N_{1}^2 N_{2}
\alpha_{1}^3 \alpha_{2}^2-128 N_{1}^2 N_{2} \alpha_{1}^4
\alpha_{3}-504 N_{1}^2 N_{2} \alpha_{1}^4 \alpha_{2}-80 N_{1}^2
N_{2} \alpha_{1}^5-64 N_{1}^2 N_{2}^2 \alpha_{2}^2 \alpha_{3}^2-432
N_{1}^2 N_{2}^2 \alpha_{2}^3 \alpha_{3}-416 N_{1}^2 N_{2}^2
\alpha_{2}^4-72 N_{1}^2 N_{2}^2 \alpha_{1} \alpha_{2}
\alpha_{3}^2-792 N_{1}^2 N_{2}^2 \alpha_{1} \alpha_{2}^2
\alpha_{3}-1200 N_{1}^2 N_{2}^2 \alpha_{1} \alpha_{2}^3-32 N_{1}^2
N_{2}^2 \alpha_{1}^2 \alpha_{3}^2-504 N_{1}^2 N_{2}^2 \alpha_{1}^2
\alpha_{2} \alpha_{3}-1280 N_{1}^2 N_{2}^2 \alpha_{1}^2
\alpha_{2}^2-144 N_{1}^2 N_{2}^2 \alpha_{1}^3 \alpha_{3}-624 N_{1}^2
N_{2}^2 \alpha_{1}^3 \alpha_{2}-128 N_{1}^2 N_{2}^2 \alpha_{1}^4-128
N_{1}^2 N_{2}^3 \alpha_{2}^2 \alpha_{3}-288 N_{1}^2 N_{2}^3
\alpha_{2}^3-144 N_{1}^2 N_{2}^3 \alpha_{1} \alpha_{2}
\alpha_{3}-528 N_{1}^2 N_{2}^3 \alpha_{1} \alpha_{2}^2-64 N_{1}^2
N_{2}^3 \alpha_{1}^2 \alpha_{3}-336 N_{1}^2 N_{2}^3 \alpha_{1}^2
\alpha_{2}-96 N_{1}^2 N_{2}^3 \alpha_{1}^3-64 N_{1}^2 N_{2}^4
\alpha_{2}^2-72 N_{1}^2 N_{2}^4 \alpha_{1} \alpha_{2}-32 N_{1}^2
N_{2}^4 \alpha_{1}^2-88 N_{1}^3 \alpha_{2}^3 \alpha_{3}^2-328
N_{1}^3 \alpha_{2}^4 \alpha_{3}-168 N_{1}^3 \alpha_{1} \alpha_{2}^2
\alpha_{3}^2-1080 N_{1}^3 \alpha_{1} \alpha_{2}^3 \alpha_{3}-104
N_{1}^3 \alpha_{1}^2 \alpha_{2} \alpha_{3}^2-1344 N_{1}^3
\alpha_{1}^2 \alpha_{2}^2 \alpha_{3}-24 N_{1}^3 \alpha_{1}^3
\alpha_{3}^2-760 N_{1}^3 \alpha_{1}^3 \alpha_{2} \alpha_{3}-168
N_{1}^3 \alpha_{1}^4 \alpha_{3}-176 N_{1}^3 N_{2} \alpha_{2}^2
\alpha_{3}^2-800 N_{1}^3 N_{2} \alpha_{2}^3 \alpha_{3}-656 N_{1}^3
N_{2} \alpha_{2}^4-240 N_{1}^3 N_{2} \alpha_{1} \alpha_{2}
\alpha_{3}^2-1760 N_{1}^3 N_{2} \alpha_{1} \alpha_{2}^2
\alpha_{3}-2160 N_{1}^3 N_{2} \alpha_{1} \alpha_{2}^3-112 N_{1}^3
N_{2} \alpha_{1}^2 \alpha_{3}^2-1376 N_{1}^3 N_{2} \alpha_{1}^2
\alpha_{2} \alpha_{3}-2688 N_{1}^3 N_{2} \alpha_{1}^2
\alpha_{2}^2-416 N_{1}^3 N_{2} \alpha_{1}^3 \alpha_{3}-1520 N_{1}^3
N_{2} \alpha_{1}^3 \alpha_{2}-336 N_{1}^3 N_{2} \alpha_{1}^4-48
N_{1}^3 N_{2}^2 \alpha_{2} \alpha_{3}^2-528 N_{1}^3 N_{2}^2
\alpha_{2}^2 \alpha_{3}-800 N_{1}^3 N_{2}^2 \alpha_{2}^3-48 N_{1}^3
N_{2}^2 \alpha_{1} \alpha_{3}^2-720 N_{1}^3 N_{2}^2 \alpha_{1}
\alpha_{2} \alpha_{3}-1760 N_{1}^3 N_{2}^2 \alpha_{1}
\alpha_{2}^2-336 N_{1}^3 N_{2}^2 \alpha_{1}^2 \alpha_{3}-1376
N_{1}^3 N_{2}^2 \alpha_{1}^2 \alpha_{2}-416 N_{1}^3 N_{2}^2
\alpha_{1}^3-96 N_{1}^3 N_{2}^3 \alpha_{2} \alpha_{3}-352 N_{1}^3
N_{2}^3 \alpha_{2}^2-96 N_{1}^3 N_{2}^3 \alpha_{1} \alpha_{3}-480
N_{1}^3 N_{2}^3 \alpha_{1} \alpha_{2}-224 N_{1}^3 N_{2}^3
\alpha_{1}^2-48 N_{1}^3 N_{2}^4 \alpha_{2}-48 N_{1}^3 N_{2}^4
\alpha_{1}-84 N_{1}^4 \alpha_{2}^2 \alpha_{3}^2-540 N_{1}^4
\alpha_{2}^3 \alpha_{3}-120 N_{1}^4 \alpha_{1} \alpha_{2}
\alpha_{3}^2-1380 N_{1}^4 \alpha_{1} \alpha_{2}^2 \alpha_{3}-52
N_{1}^4 \alpha_{1}^2 \alpha_{3}^2-1220 N_{1}^4 \alpha_{1}^2
\alpha_{2} \alpha_{3}-380 N_{1}^4 \alpha_{1}^3 \alpha_{3}-120
N_{1}^4 N_{2} \alpha_{2} \alpha_{3}^2-880 N_{1}^4 N_{2} \alpha_{2}^2
\alpha_{3}-1080 N_{1}^4 N_{2} \alpha_{2}^3-120 N_{1}^4 N_{2}
\alpha_{1} \alpha_{3}^2-1440 N_{1}^4 N_{2} \alpha_{1} \alpha_{2}
\alpha_{3}-2760 N_{1}^4 N_{2} \alpha_{1} \alpha_{2}^2-688 N_{1}^4
N_{2} \alpha_{1}^2 \alpha_{3}-2440 N_{1}^4 N_{2} \alpha_{1}^2
\alpha_{2}-760 N_{1}^4 N_{2} \alpha_{1}^3-24 N_{1}^4 N_{2}^2
\alpha_{3}^2-360 N_{1}^4 N_{2}^2 \alpha_{2} \alpha_{3}-880 N_{1}^4
N_{2}^2 \alpha_{2}^2-360 N_{1}^4 N_{2}^2 \alpha_{1} \alpha_{3}-1440
N_{1}^4 N_{2}^2 \alpha_{1} \alpha_{2}-688 N_{1}^4 N_{2}^2
\alpha_{1}^2-48 N_{1}^4 N_{2}^3 \alpha_{3}-240 N_{1}^4 N_{2}^3
\alpha_{2}-240 N_{1}^4 N_{2}^3 \alpha_{1}-24 N_{1}^4 N_{2}^4-48
N_{1}^5 \alpha_{2} \alpha_{3}^2-552 N_{1}^5 \alpha_{2}^2
\alpha_{3}-48 N_{1}^5 \alpha_{1} \alpha_{3}^2-1008 N_{1}^5
\alpha_{1} \alpha_{2} \alpha_{3}-488 N_{1}^5 \alpha_{1}^2
\alpha_{3}-48 N_{1}^5 N_{2} \alpha_{3}^2-576 N_{1}^5 N_{2}
\alpha_{2} \alpha_{3}-1104 N_{1}^5 N_{2} \alpha_{2}^2-576 N_{1}^5
N_{2} \alpha_{1} \alpha_{3}-2016 N_{1}^5 N_{2} \alpha_{1}
\alpha_{2}-976 N_{1}^5 N_{2} \alpha_{1}^2-144 N_{1}^5 N_{2}^2
\alpha_{3}-576 N_{1}^5 N_{2}^2 \alpha_{2}-576 N_{1}^5 N_{2}^2
\alpha_{1}-96 N_{1}^5 N_{2}^3-16 N_{1}^6 \alpha_{3}^2-336 N_{1}^6
\alpha_{2} \alpha_{3}-336 N_{1}^6 \alpha_{1} \alpha_{3}-192 N_{1}^6
N_{2} \alpha_{3}-672 N_{1}^6 N_{2} \alpha_{2}-672 N_{1}^6 N_{2}
\alpha_{1}-192 N_{1}^6 N_{2}^2-96 N_{1}^7 \alpha_{3}-192 N_{1}^7
N_{2}+\alpha_{2}^8 \alpha_{3}^2+6 \alpha_{1} \alpha_{2}^7
\alpha_{3}^2+15 \alpha_{1}^2 \alpha_{2}^6 \alpha_{3}^2+20
\alpha_{1}^3 \alpha_{2}^5 \alpha_{3}^2+15 \alpha_{1}^4 \alpha_{2}^4
\alpha_{3}^2+6 \alpha_{1}^5 \alpha_{2}^3 \alpha_{3}^2+\alpha_{1}^6
\alpha_{2}^2 \alpha_{3}^2+4 N_{2} \alpha_{2}^7 \alpha_{3}^2+4 N_{2}
\alpha_{2}^8 \alpha_{3}+20 N_{2} \alpha_{1} \alpha_{2}^6
\alpha_{3}^2+24 N_{2} \alpha_{1} \alpha_{2}^7 \alpha_{3}+40 N_{2}
\alpha_{1}^2 \alpha_{2}^5 \alpha_{3}^2+60 N_{2} \alpha_{1}^2
\alpha_{2}^6 \alpha_{3}+40 N_{2} \alpha_{1}^3 \alpha_{2}^4
\alpha_{3}^2+80 N_{2} \alpha_{1}^3 \alpha_{2}^5 \alpha_{3}+20 N_{2}
\alpha_{1}^4 \alpha_{2}^3 \alpha_{3}^2+60 N_{2} \alpha_{1}^4
\alpha_{2}^4 \alpha_{3}+4 N_{2} \alpha_{1}^5 \alpha_{2}^2
\alpha_{3}^2+24 N_{2} \alpha_{1}^5 \alpha_{2}^3 \alpha_{3}+4 N_{2}
\alpha_{1}^6 \alpha_{2}^2 \alpha_{3}+4 N_{2}^2 \alpha_{2}^6
\alpha_{3}^2+12 N_{2}^2 \alpha_{2}^7 \alpha_{3}+4 N_{2}^2
\alpha_{2}^8+16 N_{2}^2 \alpha_{1} \alpha_{2}^5 \alpha_{3}^2+60
N_{2}^2 \alpha_{1} \alpha_{2}^6 \alpha_{3}+24 N_{2}^2 \alpha_{1}
\alpha_{2}^7+24 N_{2}^2 \alpha_{1}^2 \alpha_{2}^4 \alpha_{3}^2+120
N_{2}^2 \alpha_{1}^2 \alpha_{2}^5 \alpha_{3}+60 N_{2}^2 \alpha_{1}^2
\alpha_{2}^6+16 N_{2}^2 \alpha_{1}^3 \alpha_{2}^3 \alpha_{3}^2+120
N_{2}^2 \alpha_{1}^3 \alpha_{2}^4 \alpha_{3}+80 N_{2}^2 \alpha_{1}^3
\alpha_{2}^5+4 N_{2}^2 \alpha_{1}^4 \alpha_{2}^2 \alpha_{3}^2+60
N_{2}^2 \alpha_{1}^4 \alpha_{2}^3 \alpha_{3}+60 N_{2}^2 \alpha_{1}^4
\alpha_{2}^4+12 N_{2}^2 \alpha_{1}^5 \alpha_{2}^2 \alpha_{3}+24
N_{2}^2 \alpha_{1}^5 \alpha_{2}^3+4 N_{2}^2 \alpha_{1}^6
\alpha_{2}^2+8 N_{2}^3 \alpha_{2}^6 \alpha_{3}+8 N_{2}^3
\alpha_{2}^7+32 N_{2}^3 \alpha_{1} \alpha_{2}^5 \alpha_{3}+40
N_{2}^3 \alpha_{1} \alpha_{2}^6+48 N_{2}^3 \alpha_{1}^2 \alpha_{2}^4
\alpha_{3}+80 N_{2}^3 \alpha_{1}^2 \alpha_{2}^5+32 N_{2}^3
\alpha_{1}^3 \alpha_{2}^3 \alpha_{3}+80 N_{2}^3 \alpha_{1}^3
\alpha_{2}^4+8 N_{2}^3 \alpha_{1}^4 \alpha_{2}^2 \alpha_{3}+40
N_{2}^3 \alpha_{1}^4 \alpha_{2}^3+8 N_{2}^3 \alpha_{1}^5
\alpha_{2}^2+4 N_{2}^4 \alpha_{2}^6+16 N_{2}^4 \alpha_{1}
\alpha_{2}^5+24 N_{2}^4 \alpha_{1}^2 \alpha_{2}^4+16 N_{2}^4
\alpha_{1}^3 \alpha_{2}^3+4 N_{2}^4 \alpha_{1}^4 \alpha_{2}^2+12
N_{1} \alpha_{2}^7 \alpha_{3}^2+64 N_{1} \alpha_{1} \alpha_{2}^6
\alpha_{3}^2+140 N_{1} \alpha_{1}^2 \alpha_{2}^5 \alpha_{3}^2+160
N_{1} \alpha_{1}^3 \alpha_{2}^4 \alpha_{3}^2+100 N_{1} \alpha_{1}^4
\alpha_{2}^3 \alpha_{3}^2+32 N_{1} \alpha_{1}^5 \alpha_{2}^2
\alpha_{3}^2+4 N_{1} \alpha_{1}^6 \alpha_{2} \alpha_{3}^2+40 N_{1}
N_{2} \alpha_{2}^6 \alpha_{3}^2+48 N_{1} N_{2} \alpha_{2}^7
\alpha_{3}+176 N_{1} N_{2} \alpha_{1} \alpha_{2}^5 \alpha_{3}^2+256
N_{1} N_{2} \alpha_{1} \alpha_{2}^6 \alpha_{3}+304 N_{1} N_{2}
\alpha_{1}^2 \alpha_{2}^4 \alpha_{3}^2+560 N_{1} N_{2} \alpha_{1}^2
\alpha_{2}^5 \alpha_{3}+256 N_{1} N_{2} \alpha_{1}^3 \alpha_{2}^3
\alpha_{3}^2+640 N_{1} N_{2} \alpha_{1}^3 \alpha_{2}^4
\alpha_{3}+104 N_{1} N_{2} \alpha_{1}^4 \alpha_{2}^2
\alpha_{3}^2+400 N_{1} N_{2} \alpha_{1}^4 \alpha_{2}^3 \alpha_{3}+16
N_{1} N_{2} \alpha_{1}^5 \alpha_{2} \alpha_{3}^2+128 N_{1} N_{2}
\alpha_{1}^5 \alpha_{2}^2 \alpha_{3}+16 N_{1} N_{2} \alpha_{1}^6
\alpha_{2} \alpha_{3}+32 N_{1} N_{2}^2 \alpha_{2}^5 \alpha_{3}^2+120
N_{1} N_{2}^2 \alpha_{2}^6 \alpha_{3}+48 N_{1} N_{2}^2
\alpha_{2}^7+112 N_{1} N_{2}^2 \alpha_{1} \alpha_{2}^4
\alpha_{3}^2+528 N_{1} N_{2}^2 \alpha_{1} \alpha_{2}^5
\alpha_{3}+256 N_{1} N_{2}^2 \alpha_{1} \alpha_{2}^6+144 N_{1}
N_{2}^2 \alpha_{1}^2 \alpha_{2}^3 \alpha_{3}^2+912 N_{1} N_{2}^2
\alpha_{1}^2 \alpha_{2}^4 \alpha_{3}+560 N_{1} N_{2}^2 \alpha_{1}^2
\alpha_{2}^5+80 N_{1} N_{2}^2 \alpha_{1}^3 \alpha_{2}^2
\alpha_{3}^2+768 N_{1} N_{2}^2 \alpha_{1}^3 \alpha_{2}^3
\alpha_{3}+640 N_{1} N_{2}^2 \alpha_{1}^3 \alpha_{2}^4+16 N_{1}
N_{2}^2 \alpha_{1}^4 \alpha_{2} \alpha_{3}^2+312 N_{1} N_{2}^2
\alpha_{1}^4 \alpha_{2}^2 \alpha_{3}+400 N_{1} N_{2}^2 \alpha_{1}^4
\alpha_{2}^3+48 N_{1} N_{2}^2 \alpha_{1}^5 \alpha_{2} \alpha_{3}+128
N_{1} N_{2}^2 \alpha_{1}^5 \alpha_{2}^2+16 N_{1} N_{2}^2
\alpha_{1}^6 \alpha_{2}+64 N_{1} N_{2}^3 \alpha_{2}^5 \alpha_{3}+80
N_{1} N_{2}^3 \alpha_{2}^6+224 N_{1} N_{2}^3 \alpha_{1} \alpha_{2}^4
\alpha_{3}+352 N_{1} N_{2}^3 \alpha_{1} \alpha_{2}^5+288 N_{1}
N_{2}^3 \alpha_{1}^2 \alpha_{2}^3 \alpha_{3}+608 N_{1} N_{2}^3
\alpha_{1}^2 \alpha_{2}^4+160 N_{1} N_{2}^3 \alpha_{1}^3
\alpha_{2}^2 \alpha_{3}+512 N_{1} N_{2}^3 \alpha_{1}^3
\alpha_{2}^3+32 N_{1} N_{2}^3 \alpha_{1}^4 \alpha_{2} \alpha_{3}+208
N_{1} N_{2}^3 \alpha_{1}^4 \alpha_{2}^2+32 N_{1} N_{2}^3
\alpha_{1}^5 \alpha_{2}+32 N_{1} N_{2}^4 \alpha_{2}^5+112 N_{1}
N_{2}^4 \alpha_{1} \alpha_{2}^4+144 N_{1} N_{2}^4 \alpha_{1}^2
\alpha_{2}^3+80 N_{1} N_{2}^4 \alpha_{1}^3 \alpha_{2}^2+16 N_{1}
N_{2}^4 \alpha_{1}^4 \alpha_{2}+64 N_{1}^2 \alpha_{2}^6
\alpha_{3}^2+300 N_{1}^2 \alpha_{1} \alpha_{2}^5 \alpha_{3}^2+564
N_{1}^2 \alpha_{1}^2 \alpha_{2}^4 \alpha_{3}^2+536 N_{1}^2
\alpha_{1}^3 \alpha_{2}^3 \alpha_{3}^2+264 N_{1}^2 \alpha_{1}^4
\alpha_{2}^2 \alpha_{3}^2+60 N_{1}^2 \alpha_{1}^5 \alpha_{2}
\alpha_{3}^2+4 N_{1}^2 \alpha_{1}^6 \alpha_{3}^2+176 N_{1}^2 N_{2}
\alpha_{2}^5 \alpha_{3}^2+256 N_{1}^2 N_{2} \alpha_{2}^6
\alpha_{3}+672 N_{1}^2 N_{2} \alpha_{1} \alpha_{2}^4
\alpha_{3}^2+1200 N_{1}^2 N_{2} \alpha_{1} \alpha_{2}^5
\alpha_{3}+976 N_{1}^2 N_{2} \alpha_{1}^2 \alpha_{2}^3
\alpha_{3}^2+2256 N_{1}^2 N_{2} \alpha_{1}^2 \alpha_{2}^4
\alpha_{3}+656 N_{1}^2 N_{2} \alpha_{1}^3 \alpha_{2}^2
\alpha_{3}^2+2144 N_{1}^2 N_{2} \alpha_{1}^3 \alpha_{2}^3
\alpha_{3}+192 N_{1}^2 N_{2} \alpha_{1}^4 \alpha_{2}
\alpha_{3}^2+1056 N_{1}^2 N_{2} \alpha_{1}^4 \alpha_{2}^2
\alpha_{3}+16 N_{1}^2 N_{2} \alpha_{1}^5 \alpha_{3}^2+240 N_{1}^2
N_{2} \alpha_{1}^5 \alpha_{2} \alpha_{3}+16 N_{1}^2 N_{2}
\alpha_{1}^6 \alpha_{3}+112 N_{1}^2 N_{2}^2 \alpha_{2}^4
\alpha_{3}^2+528 N_{1}^2 N_{2}^2 \alpha_{2}^5 \alpha_{3}+256 N_{1}^2
N_{2}^2 \alpha_{2}^6+336 N_{1}^2 N_{2}^2 \alpha_{1} \alpha_{2}^3
\alpha_{3}^2+2016 N_{1}^2 N_{2}^2 \alpha_{1} \alpha_{2}^4
\alpha_{3}+1200 N_{1}^2 N_{2}^2 \alpha_{1} \alpha_{2}^5+352 N_{1}^2
N_{2}^2 \alpha_{1}^2 \alpha_{2}^2 \alpha_{3}^2+2928 N_{1}^2 N_{2}^2
\alpha_{1}^2 \alpha_{2}^3 \alpha_{3}+2256 N_{1}^2 N_{2}^2
\alpha_{1}^2 \alpha_{2}^4+144 N_{1}^2 N_{2}^2 \alpha_{1}^3
\alpha_{2} \alpha_{3}^2+1968 N_{1}^2 N_{2}^2 \alpha_{1}^3
\alpha_{2}^2 \alpha_{3}+2144 N_{1}^2 N_{2}^2 \alpha_{1}^3
\alpha_{2}^3+16 N_{1}^2 N_{2}^2 \alpha_{1}^4 \alpha_{3}^2+576
N_{1}^2 N_{2}^2 \alpha_{1}^4 \alpha_{2} \alpha_{3}+1056 N_{1}^2
N_{2}^2 \alpha_{1}^4 \alpha_{2}^2+48 N_{1}^2 N_{2}^2 \alpha_{1}^5
\alpha_{3}+240 N_{1}^2 N_{2}^2 \alpha_{1}^5 \alpha_{2}+16 N_{1}^2
N_{2}^2 \alpha_{1}^6+224 N_{1}^2 N_{2}^3 \alpha_{2}^4 \alpha_{3}+352
N_{1}^2 N_{2}^3 \alpha_{2}^5+672 N_{1}^2 N_{2}^3 \alpha_{1}
\alpha_{2}^3 \alpha_{3}+1344 N_{1}^2 N_{2}^3 \alpha_{1}
\alpha_{2}^4+704 N_{1}^2 N_{2}^3 \alpha_{1}^2 \alpha_{2}^2
\alpha_{3}+1952 N_{1}^2 N_{2}^3 \alpha_{1}^2 \alpha_{2}^3+288
N_{1}^2 N_{2}^3 \alpha_{1}^3 \alpha_{2} \alpha_{3}+1312 N_{1}^2
N_{2}^3 \alpha_{1}^3 \alpha_{2}^2+32 N_{1}^2 N_{2}^3 \alpha_{1}^4
\alpha_{3}+384 N_{1}^2 N_{2}^3 \alpha_{1}^4 \alpha_{2}+32 N_{1}^2
N_{2}^3 \alpha_{1}^5+112 N_{1}^2 N_{2}^4 \alpha_{2}^4+336 N_{1}^2
N_{2}^4 \alpha_{1} \alpha_{2}^3+352 N_{1}^2 N_{2}^4 \alpha_{1}^2
\alpha_{2}^2+144 N_{1}^2 N_{2}^4 \alpha_{1}^3 \alpha_{2}+16 N_{1}^2
N_{2}^4 \alpha_{1}^4+200 N_{1}^3 \alpha_{2}^5 \alpha_{3}^2+808
N_{1}^3 \alpha_{1} \alpha_{2}^4 \alpha_{3}^2+1264 N_{1}^3
\alpha_{1}^2 \alpha_{2}^3 \alpha_{3}^2+944 N_{1}^3 \alpha_{1}^3
\alpha_{2}^2 \alpha_{3}^2+328 N_{1}^3 \alpha_{1}^4 \alpha_{2}
\alpha_{3}^2+40 N_{1}^3 \alpha_{1}^5 \alpha_{3}^2+448 N_{1}^3 N_{2}
\alpha_{2}^4 \alpha_{3}^2+800 N_{1}^3 N_{2} \alpha_{2}^5
\alpha_{3}+1440 N_{1}^3 N_{2} \alpha_{1} \alpha_{2}^3
\alpha_{3}^2+3232 N_{1}^3 N_{2} \alpha_{1} \alpha_{2}^4
\alpha_{3}+1664 N_{1}^3 N_{2} \alpha_{1}^2 \alpha_{2}^2
\alpha_{3}^2+5056 N_{1}^3 N_{2} \alpha_{1}^2 \alpha_{2}^3
\alpha_{3}+800 N_{1}^3 N_{2} \alpha_{1}^3 \alpha_{2}
\alpha_{3}^2+3776 N_{1}^3 N_{2} \alpha_{1}^3 \alpha_{2}^2
\alpha_{3}+128 N_{1}^3 N_{2} \alpha_{1}^4 \alpha_{3}^2+1312 N_{1}^3
N_{2} \alpha_{1}^4 \alpha_{2} \alpha_{3}+160 N_{1}^3 N_{2}
\alpha_{1}^5 \alpha_{3}+224 N_{1}^3 N_{2}^2 \alpha_{2}^3
\alpha_{3}^2+1344 N_{1}^3 N_{2}^2 \alpha_{2}^4 \alpha_{3}+800
N_{1}^3 N_{2}^2 \alpha_{2}^5+544 N_{1}^3 N_{2}^2 \alpha_{1}
\alpha_{2}^2 \alpha_{3}^2+4320 N_{1}^3 N_{2}^2 \alpha_{1}
\alpha_{2}^3 \alpha_{3}+3232 N_{1}^3 N_{2}^2 \alpha_{1}
\alpha_{2}^4+416 N_{1}^3 N_{2}^2 \alpha_{1}^2 \alpha_{2}
\alpha_{3}^2+4992 N_{1}^3 N_{2}^2 \alpha_{1}^2 \alpha_{2}^2
\alpha_{3}+5056 N_{1}^3 N_{2}^2 \alpha_{1}^2 \alpha_{2}^3+96 N_{1}^3
N_{2}^2 \alpha_{1}^3 \alpha_{3}^2+2400 N_{1}^3 N_{2}^2 \alpha_{1}^3
\alpha_{2} \alpha_{3}+3776 N_{1}^3 N_{2}^2 \alpha_{1}^3
\alpha_{2}^2+384 N_{1}^3 N_{2}^2 \alpha_{1}^4 \alpha_{3}+1312
N_{1}^3 N_{2}^2 \alpha_{1}^4 \alpha_{2}+160 N_{1}^3 N_{2}^2
\alpha_{1}^5+448 N_{1}^3 N_{2}^3 \alpha_{2}^3 \alpha_{3}+896 N_{1}^3
N_{2}^3 \alpha_{2}^4+1088 N_{1}^3 N_{2}^3 \alpha_{1} \alpha_{2}^2
\alpha_{3}+2880 N_{1}^3 N_{2}^3 \alpha_{1} \alpha_{2}^3+832 N_{1}^3
N_{2}^3 \alpha_{1}^2 \alpha_{2} \alpha_{3}+3328 N_{1}^3 N_{2}^3
\alpha_{1}^2 \alpha_{2}^2+192 N_{1}^3 N_{2}^3 \alpha_{1}^3
\alpha_{3}+1600 N_{1}^3 N_{2}^3 \alpha_{1}^3 \alpha_{2}+256 N_{1}^3
N_{2}^3 \alpha_{1}^4+224 N_{1}^3 N_{2}^4 \alpha_{2}^3+544 N_{1}^3
N_{2}^4 \alpha_{1} \alpha_{2}^2+416 N_{1}^3 N_{2}^4 \alpha_{1}^2
\alpha_{2}+96 N_{1}^3 N_{2}^4 \alpha_{1}^3+404 N_{1}^4 \alpha_{2}^4
\alpha_{3}^2+1360 N_{1}^4 \alpha_{1} \alpha_{2}^3 \alpha_{3}^2+1672
N_{1}^4 \alpha_{1}^2 \alpha_{2}^2 \alpha_{3}^2+880 N_{1}^4
\alpha_{1}^3 \alpha_{2} \alpha_{3}^2+164 N_{1}^4 \alpha_{1}^4
\alpha_{3}^2+720 N_{1}^4 N_{2} \alpha_{2}^3 \alpha_{3}^2+1616
N_{1}^4 N_{2} \alpha_{2}^4 \alpha_{3}+1840 N_{1}^4 N_{2} \alpha_{1}
\alpha_{2}^2 \alpha_{3}^2+5440 N_{1}^4 N_{2} \alpha_{1} \alpha_{2}^3
\alpha_{3}+1520 N_{1}^4 N_{2} \alpha_{1}^2 \alpha_{2}
\alpha_{3}^2+6688 N_{1}^4 N_{2} \alpha_{1}^2 \alpha_{2}^2
\alpha_{3}+400 N_{1}^4 N_{2} \alpha_{1}^3 \alpha_{3}^2+3520 N_{1}^4
N_{2} \alpha_{1}^3 \alpha_{2} \alpha_{3}+656 N_{1}^4 N_{2}
\alpha_{1}^4 \alpha_{3}+272 N_{1}^4 N_{2}^2 \alpha_{2}^2
\alpha_{3}^2+2160 N_{1}^4 N_{2}^2 \alpha_{2}^3 \alpha_{3}+1616
N_{1}^4 N_{2}^2 \alpha_{2}^4+480 N_{1}^4 N_{2}^2 \alpha_{1}
\alpha_{2} \alpha_{3}^2+5520 N_{1}^4 N_{2}^2 \alpha_{1} \alpha_{2}^2
\alpha_{3}+5440 N_{1}^4 N_{2}^2 \alpha_{1} \alpha_{2}^3+208 N_{1}^4
N_{2}^2 \alpha_{1}^2 \alpha_{3}^2+4560 N_{1}^4 N_{2}^2 \alpha_{1}^2
\alpha_{2} \alpha_{3}+6688 N_{1}^4 N_{2}^2 \alpha_{1}^2
\alpha_{2}^2+1200 N_{1}^4 N_{2}^2 \alpha_{1}^3 \alpha_{3}+3520
N_{1}^4 N_{2}^2 \alpha_{1}^3 \alpha_{2}+656 N_{1}^4 N_{2}^2
\alpha_{1}^4+544 N_{1}^4 N_{2}^3 \alpha_{2}^2 \alpha_{3}+1440
N_{1}^4 N_{2}^3 \alpha_{2}^3+960 N_{1}^4 N_{2}^3 \alpha_{1}
\alpha_{2} \alpha_{3}+3680 N_{1}^4 N_{2}^3 \alpha_{1}
\alpha_{2}^2+416 N_{1}^4 N_{2}^3 \alpha_{1}^2 \alpha_{3}+3040
N_{1}^4 N_{2}^3 \alpha_{1}^2 \alpha_{2}+800 N_{1}^4 N_{2}^3
\alpha_{1}^3+272 N_{1}^4 N_{2}^4 \alpha_{2}^2+480 N_{1}^4 N_{2}^4
\alpha_{1} \alpha_{2}+208 N_{1}^4 N_{2}^4 \alpha_{1}^2+544 N_{1}^5
\alpha_{2}^3 \alpha_{3}^2+1440 N_{1}^5 \alpha_{1} \alpha_{2}^2
\alpha_{3}^2+1248 N_{1}^5 \alpha_{1}^2 \alpha_{2} \alpha_{3}^2+352
N_{1}^5 \alpha_{1}^3 \alpha_{3}^2+736 N_{1}^5 N_{2} \alpha_{2}^2
\alpha_{3}^2+2176 N_{1}^5 N_{2} \alpha_{2}^3 \alpha_{3}+1344 N_{1}^5
N_{2} \alpha_{1} \alpha_{2} \alpha_{3}^2+5760 N_{1}^5 N_{2}
\alpha_{1} \alpha_{2}^2 \alpha_{3}+608 N_{1}^5 N_{2} \alpha_{1}^2
\alpha_{3}^2+4992 N_{1}^5 N_{2} \alpha_{1}^2 \alpha_{2}
\alpha_{3}+1408 N_{1}^5 N_{2} \alpha_{1}^3 \alpha_{3}+192 N_{1}^5
N_{2}^2 \alpha_{2} \alpha_{3}^2+2208 N_{1}^5 N_{2}^2 \alpha_{2}^2
\alpha_{3}+2176 N_{1}^5 N_{2}^2 \alpha_{2}^3+192 N_{1}^5 N_{2}^2
\alpha_{1} \alpha_{3}^2+4032 N_{1}^5 N_{2}^2 \alpha_{1} \alpha_{2}
\alpha_{3}+5760 N_{1}^5 N_{2}^2 \alpha_{1} \alpha_{2}^2+1824 N_{1}^5
N_{2}^2 \alpha_{1}^2 \alpha_{3}+4992 N_{1}^5 N_{2}^2 \alpha_{1}^2
\alpha_{2}+1408 N_{1}^5 N_{2}^2 \alpha_{1}^3+384 N_{1}^5 N_{2}^3
\alpha_{2} \alpha_{3}+1472 N_{1}^5 N_{2}^3 \alpha_{2}^2+384 N_{1}^5
N_{2}^3 \alpha_{1} \alpha_{3}+2688 N_{1}^5 N_{2}^3 \alpha_{1}
\alpha_{2}+1216 N_{1}^5 N_{2}^3 \alpha_{1}^2+192 N_{1}^5 N_{2}^4
\alpha_{2}+192 N_{1}^5 N_{2}^4 \alpha_{1}+480 N_{1}^6 \alpha_{2}^2
\alpha_{3}^2+896 N_{1}^6 \alpha_{1} \alpha_{2} \alpha_{3}^2+416
N_{1}^6 \alpha_{1}^2 \alpha_{3}^2+448 N_{1}^6 N_{2} \alpha_{2}
\alpha_{3}^2+1920 N_{1}^6 N_{2} \alpha_{2}^2 \alpha_{3}+448 N_{1}^6
N_{2} \alpha_{1} \alpha_{3}^2+3584 N_{1}^6 N_{2} \alpha_{1}
\alpha_{2} \alpha_{3}+1664 N_{1}^6 N_{2} \alpha_{1}^2 \alpha_{3}+64
N_{1}^6 N_{2}^2 \alpha_{3}^2+1344 N_{1}^6 N_{2}^2 \alpha_{2}
\alpha_{3}+1920 N_{1}^6 N_{2}^2 \alpha_{2}^2+1344 N_{1}^6 N_{2}^2
\alpha_{1} \alpha_{3}+3584 N_{1}^6 N_{2}^2 \alpha_{1}
\alpha_{2}+1664 N_{1}^6 N_{2}^2 \alpha_{1}^2+128 N_{1}^6 N_{2}^3
\alpha_{3}+896 N_{1}^6 N_{2}^3 \alpha_{2}+896 N_{1}^6 N_{2}^3
\alpha_{1}+64 N_{1}^6 N_{2}^4+256 N_{1}^7 \alpha_{2}
\alpha_{3}^2+256 N_{1}^7 \alpha_{1} \alpha_{3}^2+128 N_{1}^7 N_{2}
\alpha_{3}^2+1024 N_{1}^7 N_{2} \alpha_{2} \alpha_{3}+1024 N_{1}^7
N_{2} \alpha_{1} \alpha_{3}+384 N_{1}^7 N_{2}^2 \alpha_{3}+1024
N_{1}^7 N_{2}^2 \alpha_{2}+1024 N_{1}^7 N_{2}^2 \alpha_{1}+256
N_{1}^7 N_{2}^3+64 N_{1}^8 \alpha_{3}^2+256 N_{1}^8 N_{2}
\alpha_{3}+256 N_{1}^8 N_{2}^2 \Big)$
\end{center}

\section*{Acknowledgements}

A.Morozov is indebted for the hospitality and support to the University
of Tours, where part of this work was done.

Our work is partly supported by Russian Federal Nuclear Energy Agency,
Russian Federal Agency for Science and Innovations under contract
02.740.11.5029, by RFBR grants 07-02-00878
(A.Mir.), and 07-02-00645 (A.Mor. \& Sh.Sh.), by joint grants 09-02-90493-Ukr,
09-02-93105-CNRSL, 09-01-92440-CE, 09-02-91005-ANF and by Russian President's Grant of Support
for the Scientific Schools NSh-3035.2008.2.
The work of Sh.Shakirov is also supported in part by Moebius Contest Foundation for Young
Scientists.

\end{document}